\documentclass{aa}
\usepackage{graphicx}
\usepackage[colorlinks,allcolors=blue]{hyperref}
\bibpunct{(}{)}{;}{a}{}{,}
\usepackage{natbib}
\usepackage[varg]{txfonts}
\usepackage{xcolor}
\usepackage{mathtools}
\usepackage{subfig}
\usepackage{graphicx}
\usepackage{txfonts}
\usepackage[normalem]{ulem}
\usepackage{comment}

\def\apjl{\rm ApJL}
\def\apjs{\rm ApJS}

\def\aap{\rm A\&A}

\begin{document}

   \title{Quasi-perpendicular shocks of galaxy clusters in hybrid kinetic simulations:}

   \subtitle{The structure of the shocks}
\titlerunning{Quasi-perpendicular shocks of galaxy clusters in 2D hybrid kinetic simulations} 
\authorrunning{S.S. Boula et al.} 
   \author{S. S. Boula
          \inst{1}\fnmsep\thanks{stboula@gmail.com},
          J. Niemiec\inst{1},
          T. Amano\inst{2}
          \and
          O. Kobzar  \inst{3}
          }

   \institute{Institute of Nuclear Physics Polish Academy of Sciences, PL-31342 Krakow, Poland\\
         \and
             Department of Earth and Planetary Science, University of Tokyo,
 7-3-1 Hongo, Bunkyo-ku, Tokyo 113-0033, Japan\\
   \and
              Astronomical Observatory, Jagiellonian University, PL-30244 Krak\'{o}w, Poland\\
             }             

\date{Received September XX, XXXX; accepted March XX, XXXX}

 
  \abstract
   {Cosmic ray acceleration in galaxy clusters is still an ongoing puzzle, with relativistic electrons forming radio relics at merger shocks and emitting synchrotron radiation. These shocks are also potential sources of ultra-high-energy cosmic rays,
   gamma rays and neutrinos. 
   Our recent work focuses on electron acceleration at low Mach number merger shocks in hot intracluster medium characterized by high plasma beta. Using particle-in-cell (PIC) simulations, we previously showed that electrons are energized through stochastic shock-drift acceleration process, which is facilitated by multi-scale turbulence, including the ion-scale shock surface rippling. In the present work we perform hybrid-kinetic simulations in a range of various quasi-perpendicular foreshock conditions, including plasma beta, magnetic obliquity, and the shock Mach number.}
   {We study the ion kinetic physics, responsible for the shock structure and wave turbulence, that in turn affects the particle acceleration processes. We cover the spatial and temporal scales, which allow the development of large-scale ion turbulence modes in the system.}
   {We apply a recently developed generalized fluid-particle hybrid numerical code that can combine fluid modeling for both electrons and ions with an arbitrary number of kinetic species. We limit this model to a standard hybrid simulation configuration with kinetic ions and fluid electrons. The model utilizes the exact form of the generalized Ohm's law, allowing for arbitrary choice of mass and energy densities, as well as the charge-to-mass ratio of the kinetic species.}
   {We show that the properties of ion-driven multi-scale magnetic turbulence in merger shocks are in agreement with the ion structures observed in PIC simulations. In typical shocks with the sonic Mach number $M_s=3$, the magnetic structures and shock front density ripples grow and saturate at wavelengths reaching approximately four ion Larmor radii. Only shocks with $M_s\gtrsim 2.3$ develop ripples. At very weak shocks with $M_s\lesssim 2.3$ weak turbulence is formed downstream of the shock. 
   We observe a moderate dependence of the strength of magnetic field fluctuations on the quasi-perpendicular magnetic field obliquity. However, as the field obliquity decreases, the shock front ripples exhibit longer wavelengths. 
   Finally, we note that the steady-state structure of $M_s=3$ shocks in high-beta plasmas shows evidence that there is little difference between 2D and 3D simulations. The turbulence near the shock front seems to be a 2D-like structure in 3D simulations.}
   {}


   \maketitle
%

\section{Introduction}\label{sec:intro}
Galaxy clusters are vast collections of galaxies, hot gas, and dark matter bound together with gravitational forces. These objects often host numerous shock waves, including large-scale cluster merger shocks, generated during the collision and merging of galaxy clusters, when the gas clouds within them collide, creating shock waves that propagate through the gas and heat it to high temperatures.
Observations of radio relics, which also emit X-rays, have provided compelling evidence for the acceleration of relativistic electrons at merger shocks \citep{Brunetti14}. Such radio relics in galaxy clusters are up-to Mpc-size regions of radio emission, typically with an arc-like morphology. This emission shows strong polarization and steep integrated radio spectra ($L_{\nu}\propto \nu^{\alpha}$ with $\alpha \sim -1.1$) that steepen even further from the cluster edge towards its center \citep{vanweeren2019}. The high polarization in radio relics is attributed to the compression of the perpendicular component of preshock magnetic fields across shocks. Conversely, radio halos, formed by electrons accelerated through turbulence, exhibit notably lower polarization. Furthermore, spectral steepening within these relics primarily arises from the postshock aging of cosmic ray electrons due to synchrotron and inverse Compton losses. This allows us to gain insights into the interaction between magnetic fields and high-energy particles.

Merger shocks are found to propagate in the intracluster medium (ICM), which is hot (with temperatures reaching $10^8\,\mathrm{K}$), dilute (plasma density $10^{-3}-10^{-5}$ cm$^{-3}$) and 
weakly magnetized, with the magnetic field strength on the order of $1\mu$G. Correspondingly, ICM plasma is characterized by a high plasma beta, $\beta\gg 1$, representing the ratio of thermal to magnetic pressures. Due to high temperatures, even the most energetic merger shocks typically have low sonic Mach numbers, usually $M_\mathrm{s} \lesssim 5$.

Diffusive Shock Acceleration (DSA), also known as the first-order Fermi process, is widely considered the most plausible mechanism responsible for particle acceleration to high energies \citep[see, e.g.,][]{Bell1978a,Bell1978b,BO1978,Drury1983}. In this process, particles gain energy by repeatedly crossing the shock front while being scattered between upstream and downstream regions. The essential element of DSA is the particle injection mechanism. DSA is effective only for particles with Larmor radii larger than the shock thickness, typically a few gyroradii of thermal ions. Therefore, thermal particles should be pre-accelerated to suprathermal momenta, an order of a few ion thermal momentum. The injection is more difficult for electrons than ions due to the lower mass and smaller Larmor radii of electrons.
Thus, electron injection to DSA requires some interactions with the waves in the shock that provide their pre-acceleration. For the conditions of low-Mach-number shocks in hot ICM, such mechanisms remain not fully understood.

It has been found in numerous simulations that protons are typically more efficiently accelerated at quasi-parallel shocks, with $\theta_{B\rm n} < 45^{\circ}$, \citep[e.g.,][]{KangRyu2013, CS2014, Ryu2019}, while acceleration of electrons is more efficient at quasi-perpendicular shocks, $\theta_{B\rm n} > 45^{\circ}$, \citep{Guo14b, Guo14a, Kang2019, Ha2021, AH2022}, where $\theta_{B\rm n}$ is the angle between the shock normal and the large-scale upstream magnetic field.
The mechanisms responsible for electron injection at quasi-perpendicular ICM shocks have been extensively debated in recent years. Shock Drift Acceleration (SDA) has gained attention as a plausible mechanism for electron (pre-)acceleration in these conditions, as suggested by numerous PIC simulations \citep[e.g.,][]{KV1989, Matsukiyo2011, Guo14b, Guo14a}.
These simulations revealed that at subluminal shocks, the SDA-accelerated electrons can be reflected off the shock to propagate along the upstream mean magnetic field and excite magnetic turbulence in the foreshock region \citep{Matsukiyo2011, Guo14b, Guo14a, Kang2019}. 
This turbulence is driven by electron firehose instability (EFI) and can scatter electrons back toward the shock, enabling repeating cycles of interactions with the shock front. It has been shown, that this multi-cycle SDA process leads to the generation of supra-thermal electron spectra upstream of the shock and is more efficient in shocks with higher plasma beta. Nevertheless, the maximum electron energy that can be achieved in this process is still below the estimated injection threshold for DSA.
Recent large-scale PIC simulations of quasi-perpendicular subluminal shocks \citep{Kobzar21,Ha2021} shown that stochastic Shock Drift Acceleration \citep[SSDA,][]{Katu2019} is the primary mechanism capable of providing electron pre-acceleration up to the injection energy. During SSDA, electrons undergo stochastic pitch-angle scattering off multi-scale magnetic waves. This allows them to remain confined within the shock transition for an extended time, gaining energy through gradient drift. A crucial condition for SSDA is the presence of wideband turbulence, with wavelengths extending to the ion-scale long-wave shock rippling. Such turbulence can be driven by ion kinetic instabilities resulting from temperature anisotropies at the shock. One such instability is the Alfv'{e}n ion cyclotron (AIC) instability, frequently associated with the generation of shock rippling \citep{MM2015}.

Fully kinetic PIC simulations are often constrained by limitations in both space and time, restricting their applicability to large-scale systems due to significant computational demands. On the contrary, hybrid kinetic simulations, which combine kinetic ions and fluid electrons, can handle considerably larger macroscopic systems, which makes them particularly valuable for advancing our understanding of the ion DSA process and the ion-scale turbulence in non-relativistic shocks. A combined hybrid kinetic and test-particle model has been recently developed by \cite{TB19} to study 2D and 3D turbulent structures at low-Mach-number shocks in low-beta solar-wind plasmas and the electron acceleration therein. This approach allowed the authors to study the influence of the shock surface fluctuations on the acceleration of suprathermal electrons. 

The aim of this work is to investigate the ion-scale structure of quasi-perpendicular merger shocks with the hybrid kinetic very large-scale 2D numerical experiments, much larger than investigated previously with fully kinetic PIC simulations by \citet{Guo2019,Kobzar21,Ha2021}.  
The simulations explore a range of shock conditions, including high plasma beta, sonic Mach numbers, and subluminal shock obliquity angles. They are designed to enable us to observe the system's development of multi-scale ion-driven turbulence. The simulations are augmented with smaller-scale 3D runs.  
This analysis lays the foundation for further exploration of the role of ion-scale turbulence in electron acceleration at high-beta shocks and provides the basis for large-scale 3D simulations to study shock physics. 

The paper is organized as follows.
Section \ref{sec:2} introduces the numerical setup and methods used in the simulations. Section \ref{sec: b20} describes the 2D structure and evolution of a generic high-beta shock with the sonic Mach number $M_s=3$ and $\beta = 20$. We analyze the linear and nonlinear properties of the wave turbulence. Section \ref{sec:param} systematically explores a dependence of the shock physics on the shock and ambient plasma parameters, such as plasma beta, the sonic Mach number, and shock obliquity angle. Section \ref{App:3D} compares results from 2D and 3D simulations. Finally, Section \ref{sec:4} summarizes our findings and presents concluding remarks.


\section{Numerical Setup}
\label{sec:2}

Our studies use a new generalized fluid-particle hybrid numerical code of \cite{Amano18} that treats the collisionless plasma under the assumption of quasi-neutrality. The model generally consists of fluid ions, electrons, and an arbitrary number of kinetic species. The dynamics of fluid and kinetic populations are coupled in a self-consistent manner. The code employs the exact form of the generalized Ohm's law, which allows us to consider arbitrary mass and energy densities, as well as the charge-to-mass ratio of the kinetic species. Here, we restrict this general model to only fluid electrons and kinetic ions. Our approach thus resembles a standard hybrid model but with proper treatment of finite electron inertia effects. We assume the ion-to-electron mass ratio $m_i/m_e=100$, where  $m_i$ and $m_e$ are the ion and electron mass, respectively. We demonstrate in Appendix \ref{Ap:mass} that our results are independent of this specific choice of the mass ratio.

This work shows the results of large-scale 2D simulations supported with two smaller-scale 3D runs. The 2D simulation setup utilizes a computational grid in the $x-y$ plane. The simulation code is fully 3D, and the 2D simulations are performed with two grid points in the $z$-direction. The simulations follow all three components of particle momenta and electromagnetic fields.
An electron-ion plasma beam is injected at the right side of the computational box and flows with a bulk simulation-frame velocity $u_0$  in the $-x$-direction. After reflecting off the conductive wall at the left boundary, the beam interacts with the incoming plasma, forming a shock that travels in the positive $x$-direction at the speed $u_{sh}$, as measured in the upstream rest frame. The right $x$-boundary of the box
is open, and periodic boundary conditions are applied in transverse $y$ (and $z$) directions.

The injected plasma carries a large-scale magnetic field, $\mathbf{B_0}$, which lies in the $x-y$
plane (i.e., in-plane in 2D simulations) and is inclined at the angle $\theta_{Bn}$ 
with respect to the shock normal.
We investigate quasi-perpendicular shocks, for which $\theta_{Bn}>45^{\circ}$.
In conjunction with the magnetic field, a motional electric field, 
$\mathbf{E_0} = -[(\mathbf{u_0}/c) \times \mathbf{B_0}]$, 
is also initialized and points in the $z$-direction, $\mathbf{E_0} = E_{0z}\hat{z}$.
The results of previous PIC simulations drive the choice of the in-plane magnetic field configuration for 2D runs \cite{Kobzar21}, which showed its fidelity in reproducing 3D physics, which is also supported by our 3D simulations, see Section~\ref{App:3D}.

In the following, the velocities are normalized to the ion Alfvén speed, $u_A=B_0/\sqrt{4\pi N_{0i}m_i}$, calculated using the upstream magnetic field and the ion density, $N_{0i}$.
The ratio between Alfvén speed and the speed of light is set to $u_A/c=10^{-4}$, the adiabatic index is $\Gamma=5/3$, and the resistivity is $\eta=0$. The electrons and ions are initially in
thermal equilibrium and their temperatures are $T_i=T_e=T_0$.
The simulation-frame Mach number is defined as:
\begin{equation}
    M_{s,0}\equiv \frac{u_0}{c_s} = \frac{u_0}{\sqrt{2\Gamma k_B T_0/m_i}},
\end{equation}
where $c_s$ is the sound velocity and $k_B$ the Boltzmann constant. The shock Mach number in the shock rest frame is: 
\begin{equation}
    M_s = \frac{u_{sh}}{c_s} = M_{s,0} \left(1+\frac{1}{r(M_s)-1}\right),
\end{equation}
and the density jumps across the shock, $r(M_s)$, in the limit of weakly magnetized flows, is equal to: 
\begin{equation}
    r(M_s)=\frac{\Gamma+1}{\Gamma -1 +2/M_s^2}.
\end{equation}

The simulation parameters were selected to reproduce typical physical conditions observed in ICM shocks.
We simulate shocks
of sonic Mach number $M_s$ of total plasma beta: 
\begin{equation}
\beta= \beta_e+\beta_i= \frac{8\pi(N_{\mathrm{e}}+N_{\mathrm{i}})k_{\mathrm{B}}{T_0}}{B_0^2},    
\end{equation}
where $N_e=N_i$ is the electron number density. The plasma beta is
initially equally carried by electrons and ions, $\beta_e=\beta_i$.
If $T_e=T_i$, then the plasma beta can be expressed as:
\begin{equation}
    \beta = \frac{2 M_A^2}{\Gamma M_s^2},
\end{equation}
where $M_A=u_{sh}/u_A$ is the Alfvénic Mach number of the shock. 
Setting up the shock in the simulations with the given physical parameters of $M_s$ and $\beta$ involves the change of the normalized ion thermal velocity, $u_{th,i}/u_A=\sqrt{\beta_i/2}$, which sets up also the normalized ion sound velocity, $c_s/u_A=\sqrt{2\Gamma\beta_i/2}$, and the change
of the normalized inflow velocity, i.e., the Alfvénic Mach number of the upstream flow, $M_{A,0}=u_0/u_A$.

Our unit of space is the ion skin depth: 
\begin{equation}
 \lambda_i=\frac{c}{\sqrt{4\pi e^2N_i/m_i}},   
\end{equation}
or {the ion} gyroradius 
\begin{equation}\label{eq:Larmor}
 r_{g}\equiv \frac{u_0}{\Omega_{ci}}=\frac{u_0}{u_{sh}}M_{s}\lambda_i \sqrt{\beta\Gamma/{2}},    
\end{equation} 
where $e$ is the electric charge. The time unit is the 
ion gyrofrequency, $\Omega_{ci}={eB_0}/{m_i}c$.
The time step $\delta t=10^{-3}\Omega_{ci}^{-1}$. We resolve $\lambda_i$ with 4 cells and use $N_{ppc}=N_{\rm 0i}=128$ particles (ions) per cell.
These numerical parameters were selected based on convergence tests (see Appendix \ref{Ap:npc}).
The transverse size of the simulation box is several $r_g$ to resolve large-scale shock ripple modes. In the figures presented below, the spatial coordinates are normalized with the ion gyroradius $r_g(M_s,\beta)$ of each model, unless specified otherwise.

The list of simulation runs presented in this paper is summarized in Table~\ref{tab:parameters}.
Runs A1-A6 explore $M_s=3$ shocks in the plasma beta range $\beta=3-100$ and for $\theta_{Bn}=75^{\circ}$. The Mach number $M_s=3$ represents the typical merger shock conditions and was used in earlier PIC simulations studies \citep{Kobzar21}. The choice of $\theta_{Bn}=75^{\circ}$ enables us to directly compare the results for ion physics of hybrid-kinetic and fully-kinetic modelings. Runs B1-B6 study the quasi-perpendicular shock obliquity angle range, $\theta_{Bn}=55^{\circ}-89^{\circ}$, for $M_s=3$ and $\beta=10$. Runs C1-C6 scan the range of sonic Mach numbers, $M_s=2.0-5.5$, for $\beta=20$ and $\theta_{Bn}=75^{\circ}$. Runs D1-D2 are performed to compare the $\theta_{Bn}$-dependence for $M_s=3$ and $\beta=20$. Finally, runs E1-3D/E2-3D are the 3D simulations for $M_s=3$, $\theta_{Bn}=75^{\circ}$ and $\beta=5$ and $10$.

Investigations of merger shock structures in similar system parameter ranges have been performed earlier with 2D PIC simulations by \citet{Guo2019,Kobzar21,Ha2021}. However, due to computational constraints these simulations used moderately large transverse sizes of numerical boxes, typically $L_y\approx 21.5\lambda_i\approx (1.7-5.5) r_g $  in \citet{Guo2019}, $L_y=32\lambda_i\approx 7.8r_g$ in \citet{Kobzar21}, and $L_y =(44-62)\lambda_i\approx (3.4-5.1) r_g$ in \citet{Ha2021}, and simulation times typically $t_{max}\sim (20-32)\Omega_{ci}^{-1}$, with only two long-term runs with $t_{max}\Omega_{ci}^{-1}=50$ for $M_s=3$, $\beta=50$, and $\theta_{Bn}=73^{\circ}$ shock in \citet{Ha2021}
and $t_{max}\Omega_{ci}^{-1}=78$ for the case with $M_s=3$, $\beta=5$, and $\theta_{Bn}=75^{\circ}$ in \citet{Kobzar21}.
Compared to these studies, our hybrid simulations are very-large scale, with $L_y=(48-192)\lambda_i$, covering physical ranges from $5.9r_g$ to even $47r_g$ (see seventh and eighth columns in Table~\ref{tab:parameters}). Some of the simulations are also evolved to strongly nonlinear evolution stages, $t_{max}\gg 50\Omega_{ci}^{-1}$ (ninth column in Table~\ref{tab:parameters}). We also systematically explore a wide range of physical parameters, including plasma beta, the shock Mach number, and the magnetic field obliquity. In particular, investigations of the $\theta_{Bn}$-dependence over such a wide range of angles have not been undertaken before.

\begin{table*} 
\centering
\begin{tabular}{|c|c|c|c|c|c|c|c|c|}
\hline\hline
Run & $M_s$ & $\beta$ & $\theta_{B_n}$ & $L_x$ [$\lambda_{si}$]  & $L_x$ [$r_{g}$]& 
$L_y$ [$\lambda_{si}$] &$L_y$ [$r_{g}$] & $t_{max} [\Omega_{ci}^{-1}]$ \\
\hline
A1 & 3.0 & 3 &75 &360 & 113.9 &48& 15.2 & 26 \\
A2 & 3.0 & 5 & 75 & 360 & 88.2& 192 & 47.0& 87 \\
A3 & 3.0 & 10 & 75 & 360 & 62.4& 48 & 8.3 & 26 \\
A4 & 3.0 & 20 & 75 & 480 & 58.8 & 192 & 23.5 & 56 \\
A5 & 3.0 & 50 & 75 & 784 & 60.8 & 144 & 11.2& 62 \\
A6 & 3.0 & 100 & 75 & 384 &21.0 & 192 & 10.5& 26 \\
B1 & 3.0 & 10 & 55 & 360 & 62.4 & 48 & 8.3& 30 \\
B2 & 3.0 & 10 & 60 & 360  & 62.4 & 48 & 8.3& 30 \\
B3 & 3.0 & 10 & 65 & 360  & 62.4 & 48 & 8.3& 30 \\
B4 & 3.0 & 10 & 70 & 360  & 62.4 & 48 & 8.3& 30 \\
B5 & 3.0 & 10 & 80 & 360 & 62.4 & 48 & 8.3& 30 \\
B6 & 3.0 & 10 & 85 & 360  & 62.4 & 48 & 8.3& 30 \\
B7 & 3.0 & 10 & 89 & 360  & 62.4 & 48 & 8.3& 30 \\
C1 & 2.0 & 20 & 75 & 360 &78.4 & 96 & 20.9& 26 \\
C2 & 2.25 & 20 & 75 & 360 & 67.4& 96 &18.0 & 36 \\
C3 & 2.45 & 20 & 75 & 360 &63.0 & 96 & 16.8& 24 \\
C4 & 2.7 & 20 & 75 & 360 &50.5 & 96 &13.5 & 22 \\
C5 & 2.9 & 20 & 75 & 360 &46.0 & 96 & 12.3& 24 \\
C6 & 5.5 & 20 & 75 & 360 &22.2 & 96 &5.9 & 20 \\
D1 & 3.0 & 20 & 55 & 360 &44.1 & 192 &23.6 & 28 \\
D2 & 3.0 & 20 & 90 & 360 & 44.1& 192 & 23.6& 28 \\
E1-3D & 3.0 & 5 & 75 & 288 &70.5 & 96 &23.5 & 37 \\
E2-3D & 3.0 & 10 & 75 & 144 & 24.9 & 48 & 8.3& 40 \\
\hline
\end{tabular}
\caption{Parameters of the hybrid-kinetic simulation runs discussed in this work.}
\label{tab:parameters}
\end{table*}

\section{Evolution of the shock structure}
\label{sec: b20}

This section presents the shock structure characteristics typical of a shock with sonic Mach number $M_s=3$ and a quasi-perpendicular obliquity $\theta_{Bn}=75^{\circ}$. Our discussion is based on simulation run A4 with plasma beta $\beta=20$, which in the following is referred to as the reference run. A general shock structure is presented in Section~\ref{sec:shock}. Section~\ref{sec:linear} discusses the linear analysis of the magnetic wave modes, Section~\ref{sec:density} treats density perturbations and shock ripples, and Section~\ref{sec:nonlinear} describes the nonlinear evolution of the turbulence in the shock.

\begin{figure}
\centering
	\includegraphics[width=1.05\columnwidth]{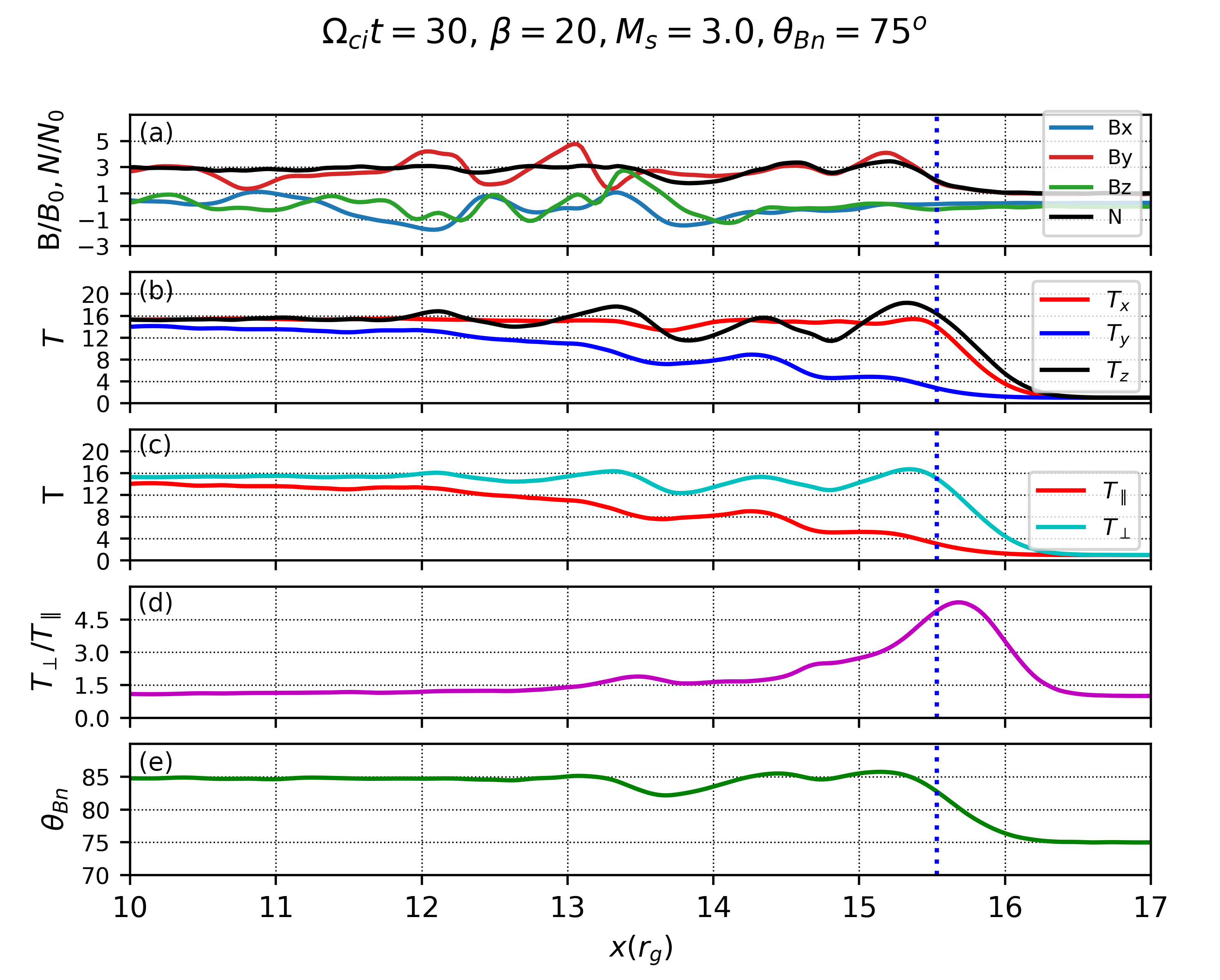}
    \caption{Structure of the reference shock with $M_s=3$, $\beta=20$, and $\theta_{Bn}=75^{\circ}$ (run A4) at time $\Omega_{ci}t=30$. Displayed are the
    $y$-averaged profiles of (a) the ion density normalized to the far upstream density $N_{0i}$ (black line), $B_x, B_y$ and $B_z$ magnetic field components normalized to their upstream values (blue, red, and green lines, respectively), (b) profiles of $T_x$, $T_y$, and $T_z$ temperature components (red, blue, and black lines, respectively), and (c) the components parallel, $T_{\parallel}$ (red line), and perpendicular, $T_{\perp}$ (cyan line), to the local magnetic field, normalized to $T_0$, (d) the $T_{\perp}/T_{\parallel}$ temperature ratio, and (e) the mean magnetic field inclination angle with respect to the shock normal, $\theta_{Bn}$. The vertical dotted line in each panel marks the shock location. The $x$ coordinate is normalized to the ion gyroradius $r_g$.}
    \label{Fig:hb20_temps}
\end{figure}

\begin{figure}
    \centering
    \includegraphics[width=\columnwidth]{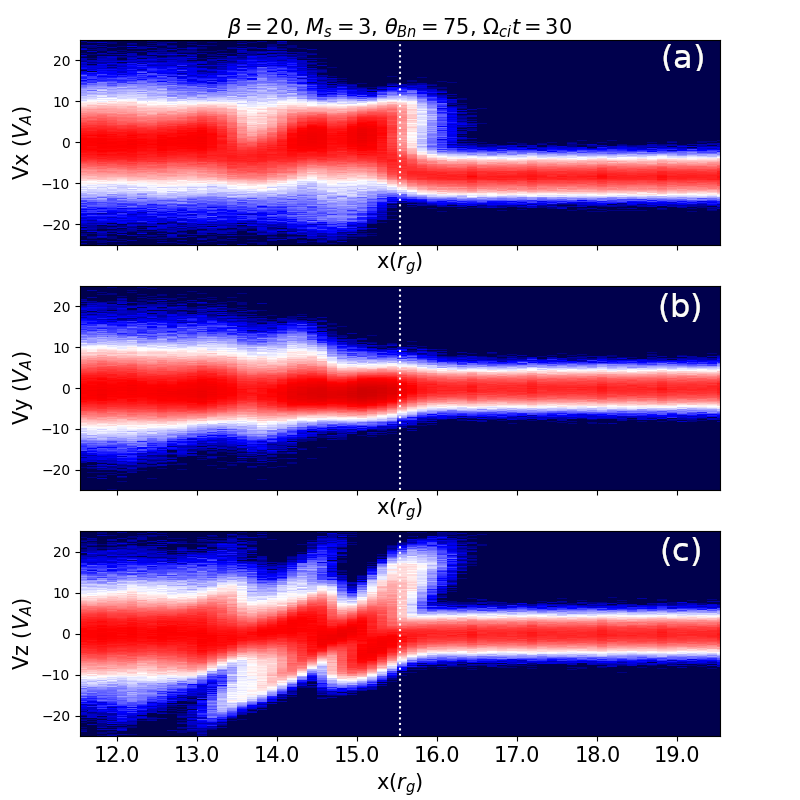}
    \caption{Ion velocity phase-space distributions, $x-\mathbf{v_i}$ ($i=x, y, z$), averaged in the $y$-direction for $M_s=3$, $\beta=20$, and $\theta_{Bn}=75^{\circ}$ shock (run A4) at time $\Omega_{ci}t=30$. The vertical dotted line marks the approximate shock location.}
    \label{Fig:hb_20_vx}
\end{figure}

\subsection{General structure of a reference shock in high-beta plasma}
\label{sec:shock}

Figure~\ref{Fig:hb20_temps} shows the profiles of various physical quantities across a well-evolved shock at time $\Omega_{ci}t=30$, averaged over the $y$-coordinate. The $x$-axis is shown in units of the ion gyroradius (Eq.~\ref{eq:Larmor}), which for the specific parameters of run A4 is $r_g\approx 4.1\lambda_i$. 
The approximate location of the shock front at $x\approx 15.5 r_g$ is shown in Figure~\ref{Fig:hb20_temps} with a vertical blue dotted line. The corresponding ion velocity phase-space is presented in Figure~\ref{Fig:hb_20_vx}.

The normalized ion density 
profile, shown with \emph{black} line in Figure~\ref{Fig:hb20_temps}a, exhibits a characteristic pattern of overshoot-undershoot oscillations past the shock front. The density compression at the double-peaked overshoot (at $x\approx 15.2r_g$ and $x\approx 14.5r_g$) reaches $N_i/N_{0}\approx 3.5$,    decreases to $N_i/N_{0}\approx 2$ in the undershoot at $x\approx 13.8 r_g$, and increases again to $N_i/N_{0}\approx 3.1$ in the second overshoot at $x\approx 13.0 r_g$.
Further downstream, the density compression gradually relaxes, eventually reaching the Rankine-Hugoniot value, approximately $r\sim3$. This observed structure is a consequence of the reflection of a portion of the incoming ions
from the shock back upstream, a phenomenon 
evidenced by the presence of ions with large positive values of $v_{xi}$ and $v_{zi}$ in the phase-space ahead of the shock at $x/r_g\approx 15.5$ 
(Fig.~\ref{Fig:hb_20_vx}a and~\ref{Fig:hb_20_vx}c). The reflected ions gyrate in the upstream magnetic field and acquire energy as they drift along the motional electric field, $E_0$. This energy increase enables the ions to surpass the potential drop across the shock and be transported downstream after a single reflection.
The reflected ions turn around back to the shock within approximately $1r_g$ and further gyrate in the downstream shock-compressed magnetic field. As the latter is nearly perpendicular to the shock normal ($\theta_{Bn}\approx 85^{\circ}$, compare Fig.~\ref{Fig:hb20_temps}e), the ion gyration occurs primarily in the $x-z$ plane, giving rise to characteristic patterns in downstream $x-v_{xi}$ and $x-v_{zi}$ phase-spaces. Such shock structures are a prevalent characteristic of quasi-perpendicular shocks \citep{Treumann2009}.

\begin{figure*}
    \centering    \includegraphics[width=1.0\linewidth]{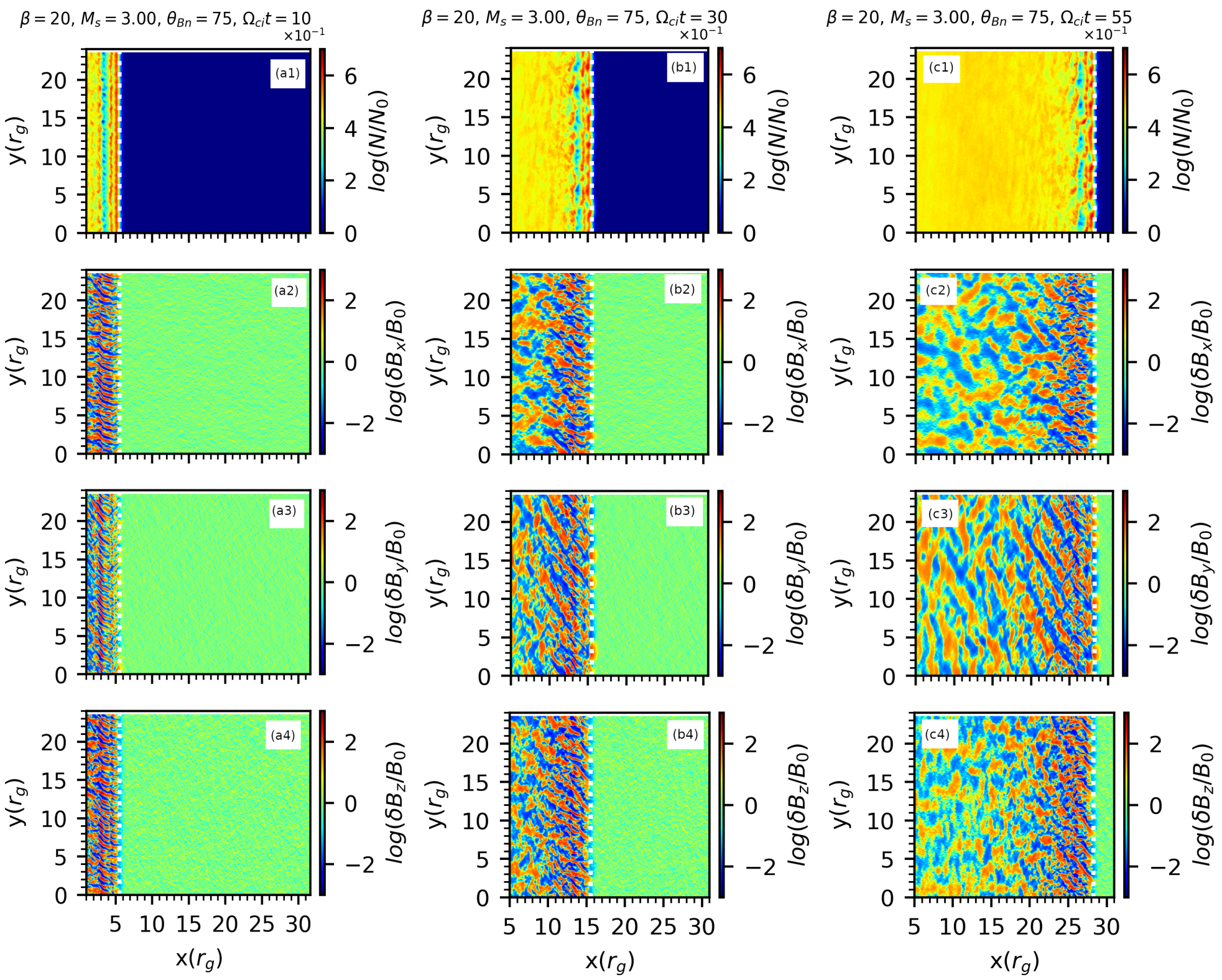} 
    \caption{Structure of the reference shock with $M_s=3$, $\beta=20$, and $\theta_{Bn}=75^{\circ}$ (run A4) at different times $\Omega_{ci}t=10$ (panels~a$^*$),  $\Omega_{ci}t=30$ (panels b$^*$), and $\Omega_{ci}t=55$ (panels c$^*$). From top to bottom, shown are the maps of the normalized ion density, $N_i/N_{0i}$ (panels~*1), and the normalized magnetic field fluctuations (see Equations~\ref{eq:dBx}-\ref{eq:dBz}), $\delta B_x$ (panels *2), $\delta B_y$ (panels *3), and $\delta B_z$ (panels *4). 
    The scaling of all distributions is logarithmic and for the magnetic fields it is sign-preserving, and, e.g., for $\delta B_{z}$ it is: $\mathrm{sgn}(\delta B_{z}) \cdot \{2+\log[\max(|\delta B_{z}|/B_{0},10^{-2})]\}$. The level of "0" on the color scale hence corresponds to $|\delta B|/B_{0} \le 10^{-2}$ \citep[see, e.g.,][]{Kobzar21}.}
    \label{Fig:hb20_maps}
\end{figure*}

The reflection and gyration of ions give rise to a marked ion temperature anisotropy 
at the shock ramp and overshoot. Figure~\ref{Fig:hb20_temps}b shows $y$-averaged profiles of the diagonal components of the ion temperature tensor, $T_x$, $T_y$, and $T_z$, normalized to $T_0$. The temperature tensor is derived from the pressure tensor, the second-order moment of the ion distribution function, calculated in the \emph{local} plasma rest frame.
In Figure~\ref{Fig:hb20_temps}c, we present the temperature components parallel, $T_{\parallel}$, and perpendicular, $T_{\perp}$, to the local magnetic field, and we show the temperature ratio, $T_{\perp}/T_{\parallel}$, in Figure~\ref{Fig:hb20_temps}d.~\footnote{To calculate the parallel and perpendicular temperature components, we apply tensor transformation rules and construct a rotation matrix around the $z$-direction, with the rotation angle corresponding to the obliquity of the \emph{local} magnetic field.
Parallel and perpendicular components are diagonal elements of the resulting tensor. Note that because off-diagonal terms in the temperature tensor are negligible and the magnetic field is nearly perpendicular to the shock normal at the overshoot and downstream, the temperature components in this region are $T_{\parallel}\approx T_y$ and $T_{\perp}\approx (T_x+T_z)/2$.}
The anisotropy is most pronounced at the shock ramp and overshoot, $T_{\perp}/T_{\parallel}\gg 1$, and its magnitude decreases with distance from the shock, while the amplitude of $T_{\perp}$ oscillations declines and $T_{\parallel}$ increases. This results from the pitch-angle scattering of ions off magnetic turbulence in the shock transition, which isotropizes the ion distribution, as one can discern in phase spaces at $x/r_g\lesssim 13$ in Figure~\ref{Fig:hb_20_vx}.

\begin{figure*}[t!]
    \centering    \includegraphics[width=1.0\linewidth]{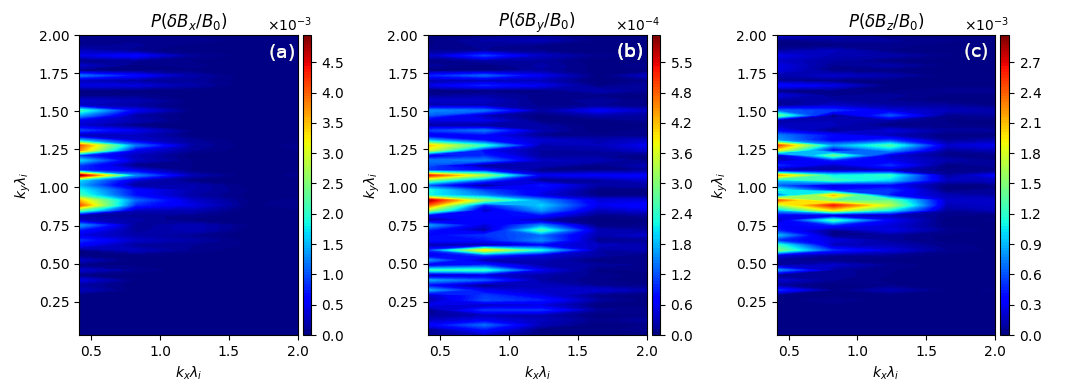}
    \caption{Fourier power spectra of the magnetic waves in the region from the shock ramp to behind the second overshoot, $1\lesssim x/r_g\lesssim 5.5$ at $\Omega_{ci}t=10$ for run A4 (compare Fig.~\ref{Fig:hb20_maps}a). Note different dynamic scales in each panel.}
    \label{Fig:fourier_t10}
\end{figure*}

The ion temperature anisotropy sources the magnetic turbulence in the shock. Figure~\ref{Fig:hb20_temps}a shows the profiles of $B_x$, $B_y$, and $B_z$ magnetic field components. The corresponding maps of the normalized ion density, $N_i/N_{0i}$, and  magnetic field fluctuations, $\delta B_x$, $\delta B_y$, $\delta B_z$, at $\Omega_{ci}t=30$ are presented in Figure~\ref{Fig:hb20_maps}b. The field fluctuations are defined with respect to the flux-frozen magnetic field, as proposed in \citet{Guo_2017}:
\begin{align}
\delta B_x=B_x-B_{0}\cos\theta_{Bn}\label{eq:dBx},\\\delta B_y=B_y -B_{0}\sin\theta_{Bn}\frac{N_i(x)}{N_{0i}}\label{eq:dBy},\\
\delta B_z=B_z\label{eq:dBz},     
\end{align}
where $N_i(x)$ is the $y$-averaged ion density. They are normalized to the upstream magnetic field, $B_0$.
The amplitudes of all components of the field fluctuations are similar and substantial enough in the shock front to induce departures from the frozen-in prediction. The latter can be discerned through deviations of the $B_y$ profile (\emph{red} line) from the density profile (a proxy for the flux-freezing prediction) in Figure~\ref{Fig:hb20_temps}a. Deviations are visible in the first overshoot and are most pronounced around the second overshoot, $x/r_g\approx 11-14$, where the amplitudes of all field components reach their maximum values.

Figures \ref{Fig:hb20_maps}a-c display snapshots of the shock structure as it evolves at times $\Omega_{ci}t=10$, 30, and 55, respectively. The ion density maps reveal corrugations in the shock transition present at the shock front, as well as in the region of the second overshoot. These shock ripples appear quite early in our simulation, approximately at time $\Omega_{ci}t\approx 6$. Their amplitude and wavelength progressively increase over time. By the time $\Omega_{ci}t=30$, the shock has transitioned into the strongly nonlinear stage. The amplitude of the shock ripples reaches $\delta N_i/N_{0}\approx 5$ and remains at this level by the end of the simulation. Strong density and the magnetic field fluctuations in the shock are confined to the region of approximately $6 r_g$ in width from the shock front (compare Figs.~\ref{Fig:hb20_temps}-\ref{Fig:hb_20_vx}). Farther downstream, the system relaxes to a weakly turbulent state. 
These characteristics of the ion-scale shock structure are consistent with results for subluminal shocks with $M_s=3$ and $\beta\sim 20$ obtained with PIC simulations \citep{Guo_2017,Ha2021}. 

\begin{figure}[b!]
    \centering    \includegraphics[width=1.0\linewidth]{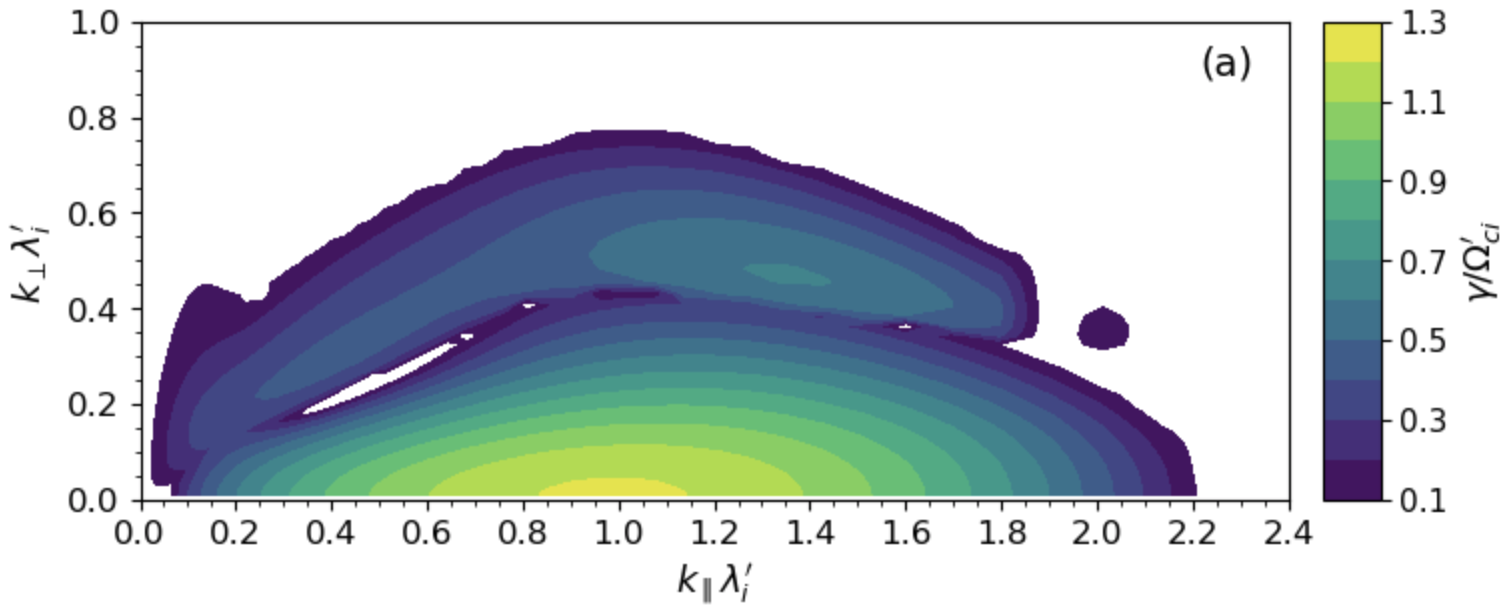}
    \includegraphics[width=1.0\linewidth]{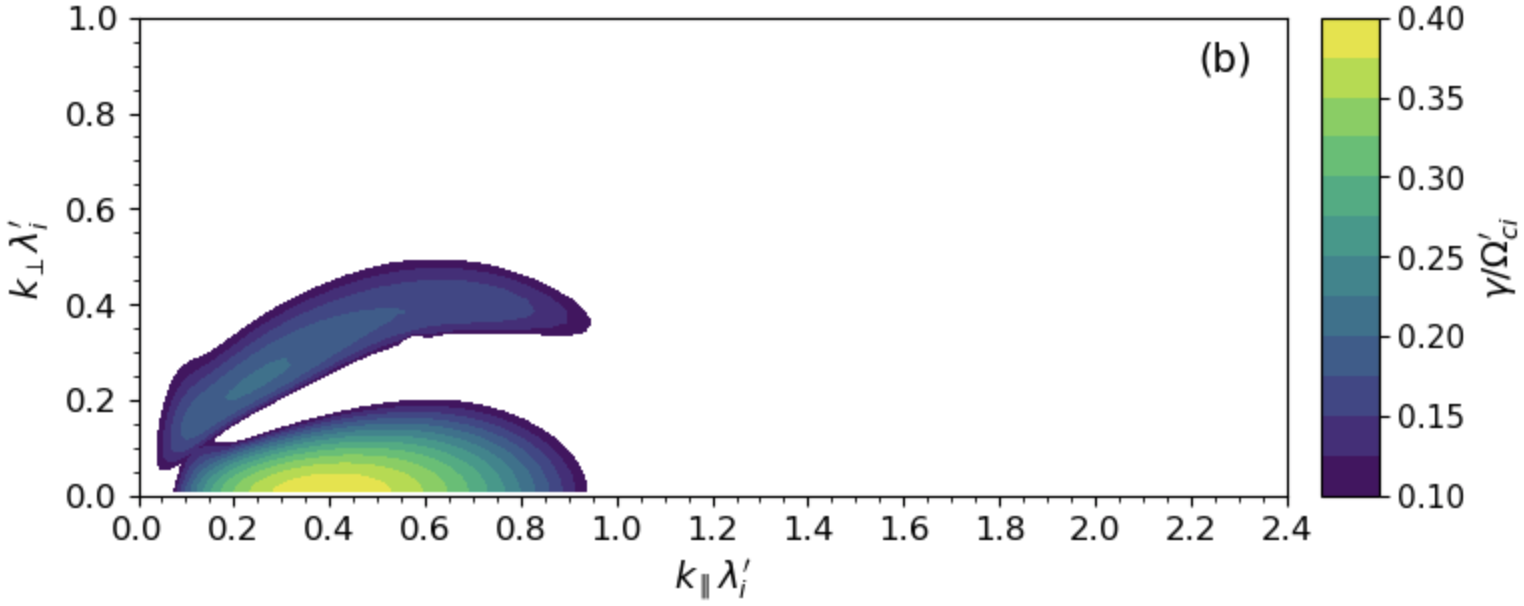}
    \caption{Linear growth rate for the AIC and mirror modes for conditions representing (a) the shock ramp and (b) the region between the shock undershoot and the second overshoot in run A4. Prime denotes quantities measured in the local plasma frame.}
    \label{Fig:linear}
\end{figure}

\subsection{Linear analysis of the wave turbulence}
\label{sec:linear}

The ion temperature anisotropy $T_{\perp}/T_{\parallel}>1$ at the shock and downstream should excite the Alfvén Ion Cyclotron (AIC) instability and the mirror instability \citep[e.g.,][]{1988WQ, 1995McKean,2003LB}. 
Figures~\ref{Fig:hb20_maps}a2-~\ref{Fig:hb20_maps}a4 show magnetic fluctuations at time $\Omega_{ci}t=10$, when the shock is at the end of the linear stage. Dominant waves in $B_x$ and $B_z$ have their wavectors primarily in the $y$-direction and so nearly aligned with the background compressed magnetic field. They are circularly polarized and move downward along the mean field direction with velocity equal to $\sim 3.7$ of the Alfven velocity in the overshoot (see below that this is compatible with the linear analysis). These waves are then consistent with AIC modes. Fluctuations in $B_y$ have oblique wave vectors that are slightly weaker, which are the characteristics of the mirror mode. The mirror modes should have zero real frequency, but in our simulation, they move in the same direction as the AIC modes. The asymmetric wave propagation in directions parallel and anti-parallel to the magnetic field has been observed except for strictly perpendicular shocks, indicating that a finite parallel drift of the reflected ions associated with obliquity plays a role. Although the linear analysis presented below assuming a symmetric ion distribution gives a reasonable agreement with the simulations in terms of wavelength, we may need to consider the finite asymmetry to understand the downward wave propagation of AIC and mirror modes. This requires a more detailed analysis of the obliquity angle dependence and is left for future study (see Section~\ref{sec:angle}).

Figure~\ref{Fig:fourier_t10} shows Fourier power spectra of the magnetic waves in the region from the shock ramp to behind the second overshoot, $1\lesssim x/r_g\lesssim 5.5$ at $\Omega_{ci}t=10$. 
The maximum wave power in $B_x$ and $B_z$ waves is at $k\sim k_y\approx 0.9/\lambda_i$, which corresponds to $\lambda\approx 7\lambda_i$. These waves are present in the undershoot and also in the first overshoot, where they co-exist with smaller-scale fluctuations with $k_y\lambda_i\approx 1.1$ and $1.3$ ($\lambda\approx 6.3\lambda_i$ and $\lambda\approx 4.4\lambda_i$, respectively), and the $(k_x,k_y)\lambda_i\approx (0.8,0.9)$ mode in $B_z$ with wavelength $\lambda\approx 5.2\lambda_i$.
The $B_y$ waves in the same region also have $k_y\lambda_i\approx 0.9, 1.1$ and $1.3$ wavevectors, though with small $k_x\lambda_i<0.7$ components. These waves are well visible in the $\delta B_y$ map in Figure~\ref{Fig:hb20_maps}a3, but not in the spectrum in Figure~\ref{Fig:fourier_t10}b, because they are much weaker than the waves present at $x/r_g\approx 3$, that dominate the Fourier power spectrum. 
The peak signal in $B_y$ waves in this second overshoot region is at $(k_x,k_y)\lambda_i\approx (0.75,0.6)$, so that $\lambda\approx 6.8\lambda_i$. Significant wave power is also in oblique modes with $(k_x,k_y)\lambda_i\approx (0.75,0.75), (0.8,0.9)$, and $(1.0,1.0)$, with the respective wavelengths of $\lambda/\lambda_i\approx 5.6, 5.2$ and $4.4$.
Finally, in the second overshoot, slightly longer wavelength modes exist in $B_x$ and $B_z$ with $k_y\lambda_i\approx 0.75$ ($\lambda\approx 8.4\lambda_i$).  

We have performed a linear dispersion analysis to identify the magnetic waves in the shock transition with the AIC and mirror modes.
The model is similar to that used in~\cite{Kim2021} and assumes a bi-Maxwellian ion velocity distribution function. Because in the shock, the ion density and velocity distributions are highly nonuniform, we adopt the numerical values of the model parameters -- the parallel ion beta $\beta_{i\parallel}\equiv\beta_{\parallel}$ and the temperature ratio, $T_{\perp}/T_{\parallel}$ -- measured at two locations across the shock: the shock ramp at $x/r_g\approx 5.5$ and the region between the undershoot and the second overshoot, $3\lesssim x/r_g\lesssim 3.5$, at time $\Omega_{ci}t=10$ (compare Fig.~\ref{Fig:hb20_maps}a). In the shock ramp, the temperature ratio is high, $T_{\perp}/T_{\parallel}\approx 7$, $\beta_{\parallel}\approx 9.3$ and plasma is weakly compressed, $r\approx 1.5$. On the other hand, in the undershoot-second overshoot region, the temperature ratio is moderate, $T_{\perp}/T_{\parallel}\approx 2.7$, $\beta_{\parallel}\approx 23$ and plasma compression $r\approx 2.7$. By calculating the linear properties in these two regions, we estimate the range of wavevectors and growth rates that should be observed in our simulation.
In Appendix~\ref{Ap:periodic} we compare the linear theory predictions for the parameters in the shock ramp with a simulation in the periodic-box. We demonstrate that linear theory results are satisfactorily reproduced.

Figure~\ref{Fig:linear} shows the linear analysis results for both regions. Parallel, $k_\parallel$, and perpendicular, $k_\perp$, wave vectors are defined with respect to the average magnetic field so that $k_\parallel$ corresponds to $k_\parallel\approx k_y$ in our simulation, and $k_\perp\approx k_x$. Note that primed units of the wave vectors and the growth rates are expressed in terms of the local parameters so that $\lambda_i^{\prime}=\lambda_i/\sqrt{r}$ and $\Omega_{ci}^{\prime}=\Omega_{ci}r$. 

The area of the strongest growth at $k_\parallel\lambda_i^{\prime}\approx 1$ and $k_\perp\approx 0$ in panel (a) for the shock ramp pertains to the AIC mode whose wavelength in the simulation units is $\lambda\approx 5.1\lambda_i$. The growth rate and the oscillation frequency of this mode are very high, $\gamma/\Omega_{ci}^{\prime}\approx 1.2$ and $\omega/\Omega_{ci}^{\prime}\approx 1.5$ 
so that the phase velocity of the AIC wave is $v_{ph}=\omega/k\approx 1.8 u_A$ (in units of the upstream Alfvén velocity). The oblique mirror mode at $(k_\parallel,k_\perp)\lambda_i^{\prime}\approx (1.35,0.45)$ has wavelength of $\lambda\approx 3.6\lambda_i$ and $\gamma/\Omega_{ci}^{\prime}\approx 0.67$. 
The AIC mode in the undershoot-second overshoot region in panel (b) peaks at much smaller linear growth of $\gamma/\Omega_{ci}^{\prime}\approx 0.4$ and much smaller  $k_\parallel\lambda_i^{\prime}\approx 0.4$, which corresponds to $\lambda\approx 9.6\lambda_i$. With the real frequency of $\omega/\Omega_{ci}^{\prime}\approx 0.8$, the phase velocity of this mode is $v_{ph}\approx 3.3 u_A$. A~weakly growing mirror mode with $\gamma/\Omega_{ci}^{\prime}\approx 0.2$ also has a longer wavelength of $\lambda\approx 7.6\lambda_i$, corresponding to $(k_\parallel,k_\perp)\lambda_i^{\prime}\approx (0.35,0.35)$. 
We conclude that the predicted range of the AIC and mirror modes wavelengths is consistent with waves with $\lambda/\lambda_i\approx 4-8$ observed in the shock transition. The estimated growth time scales, $\tau\equiv 1/\gamma$, of the AIC modes, $\tau_{AIC}\approx (0.5-0.9)/\Omega_{ci}$, and the mirror modes, $\tau_{mirror}\approx (1.0-1.9)/\Omega_{ci}$, are also in good agreement with the fact that both modes develop very early in our simulation. 

We note finally that the ion-scale magnetic field turbulence structure observed in our hybrid simulations agrees with the fully kinetic PIC simulations performed for the same physical parameters. A detailed comparison is provided in Appendix~\ref{Ap:PIC_hybrid}. The alignment found in the results derived from the hybrid-kinetic method and the PIC simulations confirms the reliability of hybrid simulations for investigating ion-scale shock physics.  

\subsection{Plasma density perturbations and shock ripples}
\label{sec:density}

The lack of electron-scale waves in hybrid simulations in the linear stage enables us to witness the initial evolution of plasma density fluctuations in the shock front (compare Appendix~\ref{Ap:PIC_hybrid}).
One can see in Figure \ref{Fig:hb20_maps}a and also in Figure~\ref{Fig:hybridpicb5b20}d, that perturbations in ion density on top of the shock plasma compression appear very early in the reference shock simulation. 
This is also true for other plasma betas investigated here for shocks with $M_s=3$ (compare, e.g., Fig.~\ref{Fig:hybridpicb5b20}b for $\beta=5$).
Previous work suggested that shock rippling is associated with AIC instability. 
However, since this instability is purely electromagnetic and transverse, generating density compressions must involve magnetic perturbations with components parallel to the mean magnetic field, such as the oblique mirror modes. The characteristics of the ripple wave structure and evolution observed in our simulations are in agreement with other studies suggesting that density fluctuations at the shock front are the effects of the nonlinear evolution of AIC and mirror modes, along with nonlinear couplings between these modes \citep[e.g.,][]{1988WQ,Lowe1999,Shoji2009,Kim2021}.

Earlier PIC simulations of shocks in high-beta plasmas by \cite{Kobzar21} and \cite{Ha2021} reported much slower growth rates and much longer wavelengths of the initial shock front ripple modes than observed in our hybrid simulation. 
For example, \cite{Kobzar21} for $M_s=3$ and $\beta=5$ shock observed the ripples to emerge at $\Omega_{ci}t\approx 25$ with wavelength $\lambda_{\text{ripple}} \approx 16\lambda_i$, in agreement with their linear dispersion analysis that predicted the peak growth rate of the AIC modes of $\gamma/\Omega_{ci}\approx 0.076$.
As we demonstrate in Section~\ref{sec:linear}, in our hybrid reference run, the initial growth rates are much faster, and AIC mode wavelengths are much shorter than these values, which is also in line with our linear analysis that approximates numerical values at the overshoot. Moreover, we show in Appendix~\ref{Ap:PIC_hybrid} that results of A4 and A2 simulations reproduce the early ion-scale structures of the magnetic field fluctuations and plasma density in PIC simulations. 
However, note that there is no disagreement between these findings.
The linear analysis by \cite{Kobzar21} is based on a different model that takes into account the estimates of the number of ions transmitted and reflected at the shock, and the temperature ratio $T_{\perp}/T_{\parallel}=4.7$ at an already evolved shock.
On the other hand, the linear analysis model used by \cite{Kim2021} is similar to ours but takes the parameters from shock simulations averaged over an extended shock transition zone, e.g., $T_{\perp}/T_{\parallel}=2$ for $M_s=3$ shocks in $\beta=5-100$ plasmas. 
Consequently, these linear estimates provide growth rates and wavelengths matching the shock conditions at later nonlinear evolution stages, at which our hybrid simulations show comparable turbulence characteristics.

\begin{figure}[b!]
	\centering
 \includegraphics[width=0.9\columnwidth]{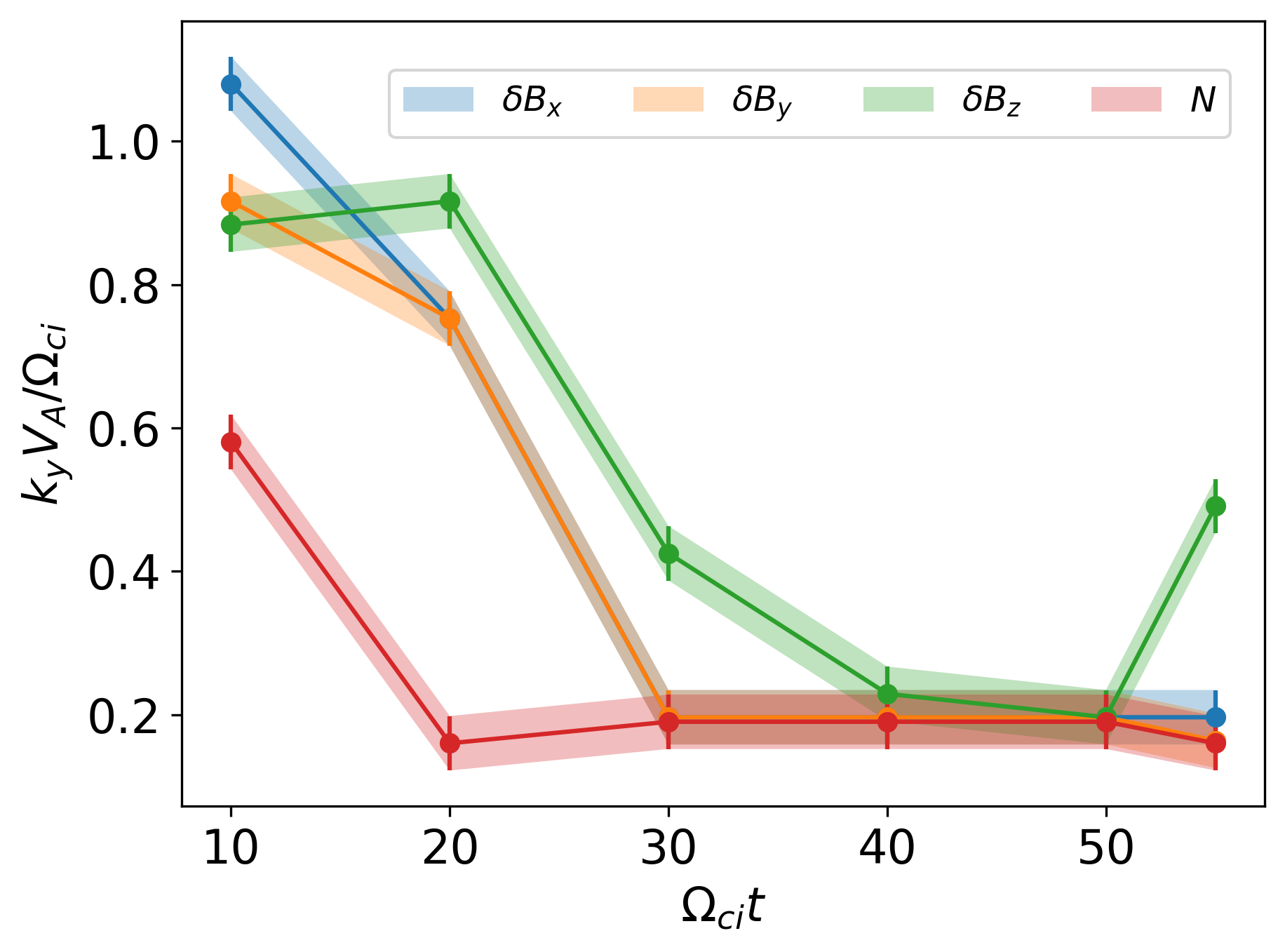}
    \caption{Temporal evolution of the dominant $k_y$ wavevector component of the magnetic field and ion density fluctuations in the shock front for the reference simulation with $M_s=3$, $\beta=20$, and $\theta_{Bn}=75^{\circ}$ (run A4). The error bars reflect resolution in the Fourier space (see text).}
    \label{Fig:wavevector}
\end{figure}

\subsection{Nonlinear evolution of the shock structure}
\label{sec:nonlinear}

As noted above, 
in our reference simulation, we observe a transition of the shock system into a nonlinear phase already at $\Omega_{ci}t\approx 6$. One can see in Figure~\ref{Fig:hb20_maps} that with time, the magnetic field and plasma density fluctuations in the shock front grow in wavelength, and large-scale shock ripples develop. 
Figure~\ref{Fig:wavevector} illustrates the temporal evolution of the magnetic field and density turbulence in the region of the forward peak of the first overshoot. Shown are the values of the $k_y$ wavevector component that is approximately aligned with the compressed magnetic field direction for individual components of the magnetic field as well as the ion density.
Although the wave power spectra of turbulence reveal strong signals at multiple distinct wavevectors, we display only the values of the dominant $k_y$ (compare Fig.~\ref{Fig:fourier_t10} for magnetic field data at $\Omega_{ci}t=10$).

One can note that with time the spectrum of magnetic AIC (primarily $B_x$ and $B_z$ components) and mirror (mainly $B_y$ component) waves shifts to lower wave numbers. Such energy transfer from higher to lower wave numbers is due to nonlinear effects, such as the parametric decay instability \citep{Shoji2009}.  
As time progresses beyond $\Omega_{ci}t\approx 30$, the field components exhibit saturation at $k_y\approx 0.2$, corresponding to a wavelength of approximately $30\lambda_i$. 

Similar evolution from short to long wavelengths followed by saturation is observed for density fluctuations. Past $\Omega_{ci}t\approx 20$ the ripple wavelength is $\lambda_{\text{ripple}} \approx 30\lambda_i$. The ripple structure is neither steady nor regular, though, as at locations along the shock surface, smaller ripples appear at times in addition to the larger-scale ones (see Fig.~\ref{Fig:hb20_maps}b-c. However, the structures do not evolve toward even longer wavelengths by the end of our reference run, A4 at $\Omega_{ci}t=56$. 
On the other hand, weak downstream turbulence beyond the second overshoot remains at a quasi-steady state at times $\Omega_{ci}t\gtrsim 30$.
Note, that we observe similar behavior for $M_s=3$ shock propagating in $\beta=5$ plasma (run A2), which we follow up to $\Omega_{ci}t=87$. 
To the best of our knowledge, these characteristics of the nonlinear evolution and saturation are reported here for the first time in the context of weak shocks in high-beta plasmas. 
This is facilitated in the hybrid approach, which enables extended, large-scale simulations of shocks, a capability beyond the current constraints of PIC simulations.

\begin{figure}
 \centering
 \includegraphics[width=0.76\linewidth]{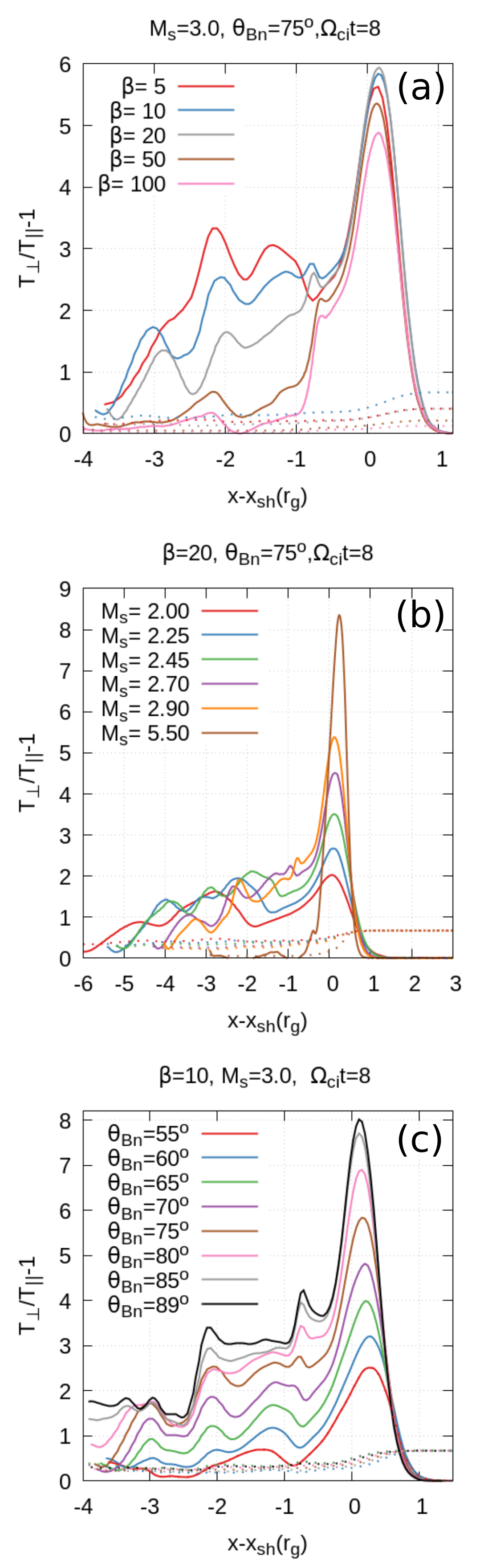}
 \caption{The $y$-averaged profiles of the ion temperature anisotropy (solid lines), and the marginal stability condition of Equation~\ref{eq:gary} (dotted lines) at time $\Omega_{ci}t=8$ for shocks with different plasma $\beta$ and $M_s=3,~\theta_{Bn}=75^{\circ}$ (runs A2-A6, panel a), different sonic Mach numbers and $\beta=20,~\theta_{Bn}=75^{\circ}$ (runs C1-C6, panel b), and a range of obliquity angles for $M_s=3,~ \beta=10$~(runs A3 and B1-B7, panel~c). The $x$ coordinate is normalized to the ion gyroradius $r_g(M_s,\beta)$ of each model.}
\label{Fig:comp_temp}
\end{figure}

\section{Parameter dependence}
\label{sec:param}

In this section we discuss the dependence of the results obtained for the reference run A4 with $M_s=3$, $\beta=20$, and $\theta_{Bn}=75^{\circ}$ on the shock parameters: plasma beta in the range $\beta=5-100$ in Section~\ref{sec:beta}, the shock Mach number in the range $M_s=2-5.5$ in Section~\ref{sec:mach}, and the range of the upstream magnetic field inclination angle, $\theta_{B_n}=55^\circ-90^\circ$, in Section~\ref{sec:angle}.

For all shocks analyzed here, the motions of shock-reflected ions give rise to the ion temperature anisotropies that catalyze the development of the ion cyclotron and mirror modes. These waves scatter the ions in pitch angle and reduce their anisotropy, bringing it closer to the upper limit provided by the marginal stability condition discussed in \cite{Gary1997}:
\begin{equation}\label{eq:gary}
\frac{T_{\perp}}{T_{\parallel}}-1\simeq \frac{1.6}{\beta_{\parallel}^{0.72}},  
\end{equation}
where $\beta_{\parallel}=8\pi N_{\mathrm{i}}k_{\mathrm{B}}{T_\parallel}/B^2$ 
is the proton plasma beta at a given location, calculated using the proton temperature in the parallel direction.
Figure~\ref{Fig:comp_temp} shows average profiles of the ion temperature anisotropy (solid lines) for runs A2-A6 for shocks with different plasma $\beta$ and $M_s=3,~\theta_{Bn}=75^{\circ}$ (panel a), runs C1-C6 for different shock sonic Mach numbers and $\beta=20,~\theta_{Bn}=75^{\circ}$ (panel b), and runs A3 and B1-B7 for a range of obliquity angles $\theta_{B_n}=55^\circ-89^\circ$ for shocks with $M_s=3$ and $\beta=10$~(panel c). 
According to Equation~\ref{eq:gary}, marginal stability limits are shown in these plots with dotted lines. The profiles are presented at an early stage of the shock evolution at time $\Omega_{ci}t=8$.   

\begin{figure*}[t!]
    \centering
    \includegraphics[width=\linewidth]{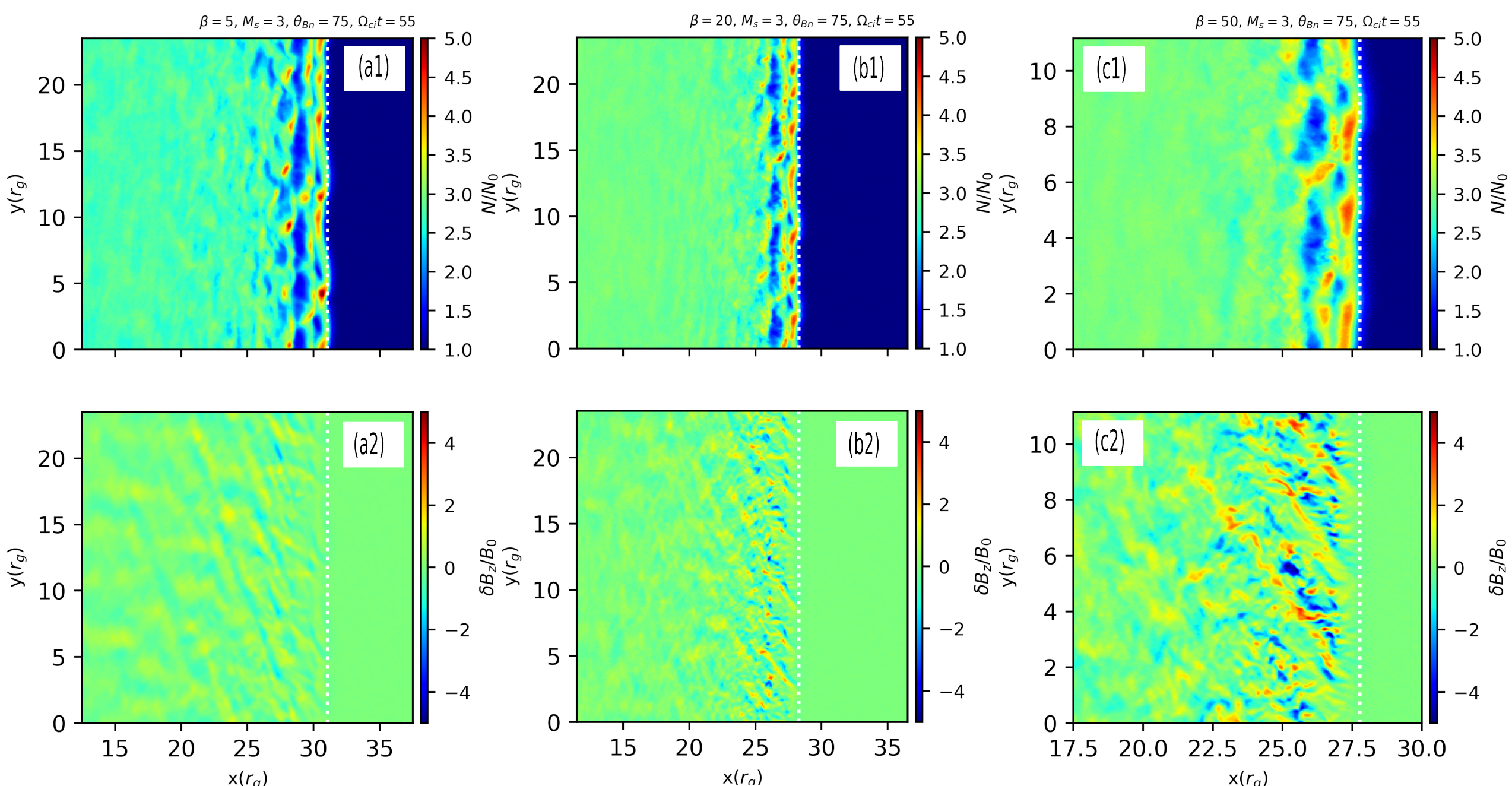}
    \caption{Structure of shocks with $M_s=3$ and $\theta_{Bn}=75^{\circ}$ at $\Omega_{ci}t=55$ for different values of plasma beta, $\beta=5,~20,~50$, runs A2, A4, and A5 in panels a-c, respectively.
    Upper panels (*1) show distributions of normalized ion density, and lower panels (*2) the normalized $B_z$ magnetic field fluctuations. The scaling is linear (compare Fig.~\ref{Fig:hb20_maps}c for the case with $\beta=20$, run A4). White dotted lines depict the approximate location of the shock, $x_{\rm sh}$, with respect to which distance is calculated in Figures~\ref{fig:db_betas} and \ref{fig:db_betas55}.}
    \label{Fig:beta_maps}
\end{figure*}

\begin{figure}
	\centering
 \includegraphics[width=0.9\columnwidth, trim={0 3.5cm 0 3.5cm}, clip]{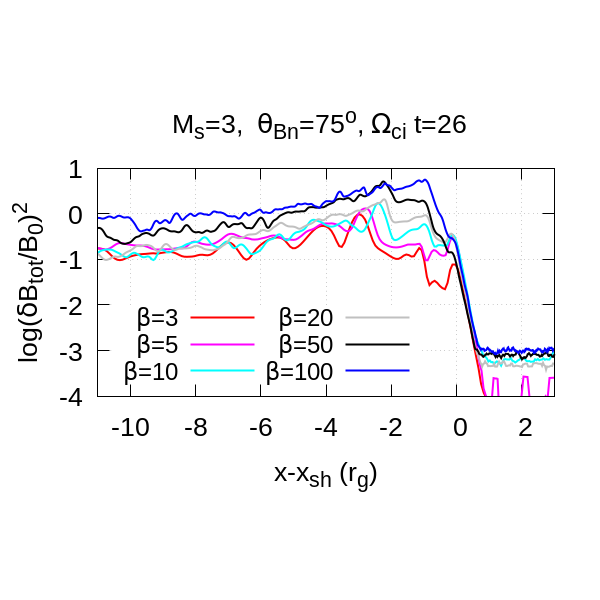}
     \caption{The $y$-averaged profiles at time $\Omega_{ci}t=26$ of the normalized total energy density of the magnetic field fluctuations across the shocks with $M_s=3$ and $\theta_{Bn}=75^{\circ}$ and plasma beta in the range $\beta=3-100$, runs A1-A6. The $x$ coordinate is normalized to the ion gyroradius $r_g$ and measured relative to the shock location $x_{\rm sh}$.}
     \label{fig:db_betas}
\end{figure}

\begin{figure}
	\centering\includegraphics[width=1.0\columnwidth]{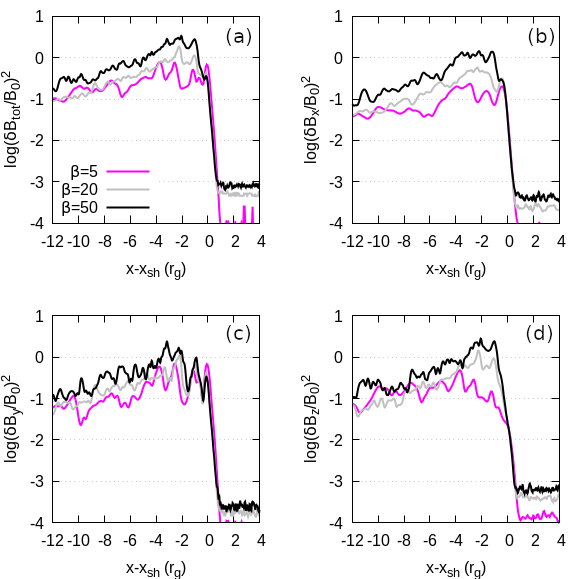}
     \caption{The $y$-averaged profiles at time $\Omega_{ci}t=55$ of the normalized energy density of the magnetic field fluctuations (total and in $B_x$, $B_y$, and $B_z$ components) across shocks with $M_s=3$ and $\theta_{Bn}=75^{\circ}$ and plasma $\beta=5$, 20, and 50, runs A2, A4, A5.}
     \label{fig:db_betas55}
\end{figure}

\subsection{Dependence on plasma $\beta$}
\label{sec:beta}

Figure~\ref{Fig:beta_maps} shows maps of the ion density and $B_z$ magnetic field fluctuations for the selected cases of $\beta=5$, 20, and 50, for shock Mach number $M_s=3$, and $\theta_{B_n}=75^\circ$. 
One can see in Figure~\ref{Fig:comp_temp}a, that the amplitude of the temperature anisotropy peak at the shock front remains relatively unchanged, regardless of the plasma beta.
At the same time, higher $\beta$ values are associated with lower marginal stability limits, resulting in faster growth rates of ion-scale instabilities. Consequently, shocks in higher-$\beta$ plasmas initially generate more intense turbulence that provides more efficient ion scattering, leading to a quicker decline in anisotropy behind the shock. 
Therefore, in high-$\beta$ shocks, turbulence attains its peak and becomes localized closer to the shock front (Fig.~\ref{Fig:beta_maps}b-c for $\beta=20$ and 50). Conversely, lower-$\beta$ shocks sustain a finite level of ion isotropy further downstream. The turbulence exists in a more extended downstream region,  where the magnetic energy also reaches its maximum (Fig.~\ref{Fig:beta_maps}a for $\beta=5$). 

The relationship between the plasma beta and the location of the peak magnetic energy is most evident in Figures~\ref{fig:db_betas} and~\ref{fig:db_betas55} that present profiles of the magnetic field energy density at times $\Omega_{ci}t=26$ and $\Omega_{ci}t=55$, respectively. The maximum magnetic energy density is observed approximately $1 r_g$ behind the shock with $\beta=100$, about $2 r_g$ for $\beta=50$ and 20, to around $3 r_g$ for $\beta=5$. Lower-$\beta$ shocks develop weaker turbulence, Figure~\ref{Fig:beta_maps}.
By the early nonlinear simulation phase at $\Omega_{ci}t=26$, a substantial difference in the total magnetic energy, spanning nearly one order of magnitude, is observed when the plasma beta varies by a factor of $\sim 30$, Figure~\ref{fig:db_betas}. The amplitude levels do not change in the strongly nonlinear stage; compare Figure~\ref{fig:db_betas55}a. 
From the early stage of the evolution, the magnetic turbulence is dominated by the $B_z$ field component, Figure~\ref{fig:db_betas55}b-d. The amplitude difference between $B_z$ and $B_x$ fluctuations is an artifact of the 2D geometry, 
while the smaller strength of $B_y$ waves is attributed to the lower growth rate of the mirror mode instability.
These findings are consistent with PIC simulation results \citep{Guo_2017,Guo2019}. 

Runs A2 and A5, have been followed to late times $\Omega_{ci}t\geq 62$. As in the reference run, we observe a saturation of the ripple growth. The nonlinear wavelengths of the ripples are $\lambda_{\text{ripple}} \approx 16\lambda_i$ and $\lambda_{\text{ripple}} \approx 48\lambda_i$ for $\beta=5$ and $\beta=50$, respectively. 
In units of the ion skindepth, the wavelengths thus grow with plasma beta, as expected. However, note that in units of the ion gyroradius, the nonlinear ripple wavelength is approximately independent of beta, and for the three cases with $\beta=5, 20$ and 50 it is $\lambda_{\text{ripple}} \approx 3.7-3.9 r_g$.

\begin{figure*}
    \centering
       \includegraphics[width=\linewidth]{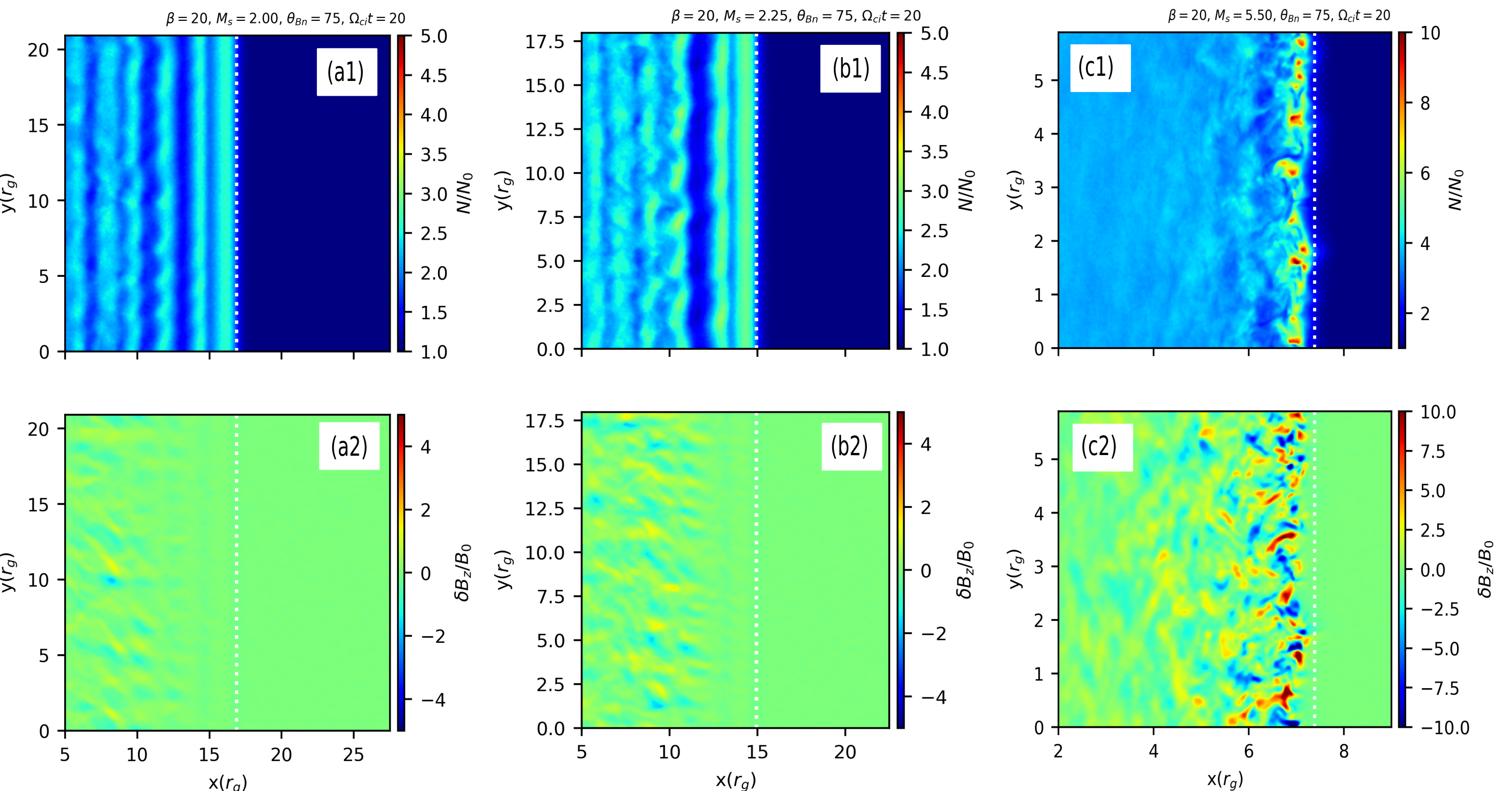}
    \caption{Structure of shock with $\beta=20$ and $\theta_{Bn}=75^{\circ}$ at time $\Omega_{ci}t=20$ for different values of the sonic Mach number, $M_s=2.00, ~2.25, ~ 5.50$, 
    runs C1, C2, and C6 in panels a-c, respectively (compare Fig.~\ref{fig:db_betas55}).
    }
    \label{Fig:ms_maps}
\end{figure*}

\begin{figure}
	\centering\includegraphics[width=1.0\columnwidth, trim={0 1cm 0 1cm}, clip]{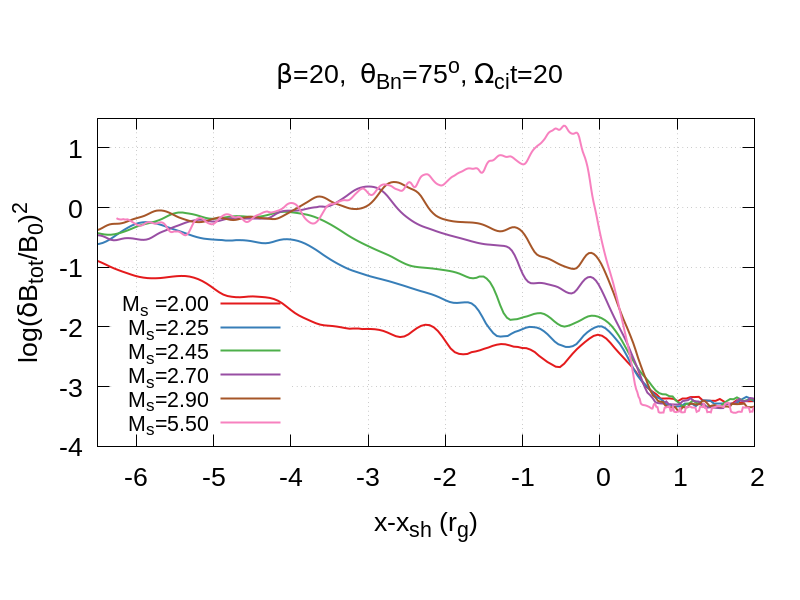}
     \caption{The $y$-averaged profiles at time $\Omega_{ci}t=20$ of the normalized total energy density of the magnetic field fluctuations across shocks with $\beta=20$, $\theta_{Bn}=75^{\circ}$ and Mach numbers in a range $M_s=2-5.5$, runs C1-C6 (compare Fig.~\ref{fig:db_betas}).}
     \label{fig:db_mcr}
\end{figure}

\begin{figure*}
    \centering
     \includegraphics[width=\linewidth]{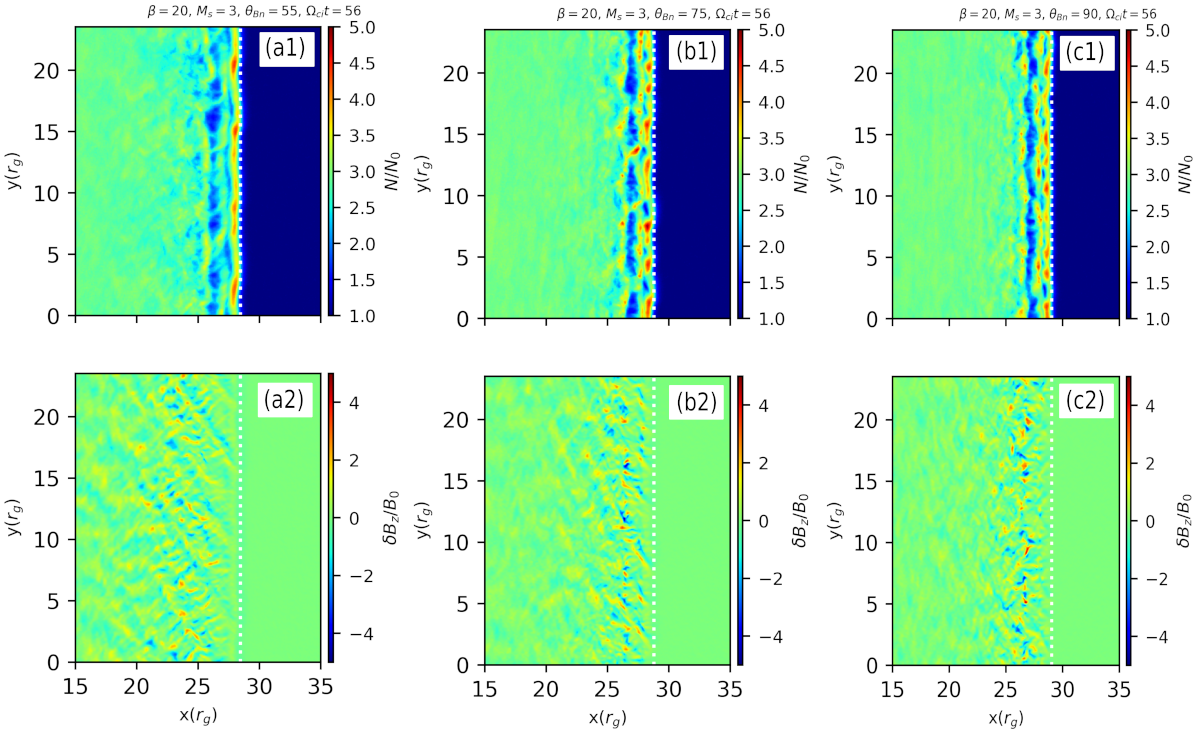}
    \caption{Structure of shocks with $\beta=20$ and $M_s=3$ at time $\Omega_{ci}t=56$ for different values of the obliquity angle, $\theta_{Bn}=55^\circ,~75^\circ$ and $~90^\circ$, 
    runs D1, A4, and D2 in panels a-c, respectively (compare Fig.~\ref{fig:db_betas55}).}
    \label{Fig:theta_comp}
\end{figure*}

\subsection{Dependence on the sonic Mach number $M_s$}
\label{sec:mach}

Figure~\ref{Fig:ms_maps} shows maps of the ion density and $B_z$ magnetic field fluctuations for the example cases of shocks with the shock Mach number $M_s=2$, 2.25, and 5.5 in
$\beta=20$ plasmas with $\theta_{B_n}=75^\circ$. 
Figure~\ref{fig:db_mcr} presents the profiles of the total energy density of the magnetic field turbulence for all simulation runs C for shocks in the Mach number range $M_s=2-5.5$.
One can see in Figure~\ref{Fig:comp_temp}b that the amplitude of the temperature anisotropy peak at the shock front is a factor of approximately 4 larger for a strong shock with $M_s=5.5$ compared to a weak shock with $M_s=2.0$. At the same time, the marginal stability limits at the shock are similar in each case. 
In the case of the $M_s=5.5$ shock, the anisotropy drops right at the shock front, whereas for lower-$M_s$ shocks, it decreases in gradually wider downstream regions. In effect, since larger temperature anisotropies lead to faster growth of instabilities and stronger turbulence, in a high-$M_s=5.5$ shock, the turbulence attains its peak within less than $1 r_g$ downstream from 
the shock front, Fig.~\ref{Fig:ms_maps}c and Fig.~\ref{fig:db_mcr} (magenta line).
With decreasing Mach number, the waves grow slower and hence the magnetic energy density reaches maximum progressively farther from the shock front, from about $2.5 r_g$ for $M_s=2.9$ shock, through approximately $3 r_g$, $4 r_g$, and $6 r_g$, respectively for shocks with $M_s=2.7, 2.45$, and 2.25. At the same time, the turbulence extends in wider downstream regions and becomes weaker with lower $M_s$. 
One can see in Fig.~\ref{fig:db_mcr}, that the amplitude of the magnetic field is two orders of magnitude smaller for the $M_s=2$ shock compared to the $M_s\approx 3$ shock and more than three orders of magnitude smaller compared to the strong $M_s=5$ shock (compare also Fig.~\ref{Fig:ms_maps}). 

One can see in the upper panels of Figure~\ref{Fig:ms_maps} and in  Figure~\ref{Fig:hb20_maps}, that higher-$M_s$ shocks, $M_s=3, 5.5$, develop distinct shock front ripples. Density corrugations in the shock front, in particular in the backward peak of the first overshoot, are also visible in the $M_s=2.25$ shock (Fig.~\ref{Fig:ms_maps}b1), whereas the $M_s=2$ shock surface is smooth and the ion-scale turbulence is built only in the region of the second overshoot ($\sim 5 r_g$ behind the shock) and further downstream.

Based on similar observations from PIC simulations, \cite{Ha2021} suggested that shock surface ripples appear only in shocks with the sonic Mach number above the critical Mach number $M_{\rm AIC}^*\approx 2.3$. At and above this $M_s$ value, the temperature anisotropy averaged over a fixed width in the shock transition, encompassing the first and the second overshoots, significantly exceeded the marginal stability condition, as indicated on the right-hand side of Equation~\ref{eq:gary}.
This coincided with an enhanced ion reflection efficiency from the shock at shocks with $M_s\gtrsim 2.3$.
At shocks with $M_s<M_{\rm AIC}^*$, the temperature anisotropy was $T_{\perp}/T_{\parallel}\approx 1$ on average, and hence below the stability limit.
In effect, the AIC instability was not excited in these shocks.
The sonic Mach number at which the shock ripples appear in our hybrid simulations is consistent with these findings. However, we note that in all our simulation runs C, the temperature anisotropy is markedly above the marginal stability condition (e.g., $T_{\perp}/T_{\parallel}\gtrsim 2$ within $\sim 5 r_g$ of the $M_s=2$ shock front) and relaxes via pitch angle scattering only far downstream of the second overshoot for very weak shocks (compare Fig.~\ref{Fig:comp_temp}b).
Because the ion temperature anisotropy in shocks with $M_s\lesssim 2.3$ is small, the resulting growth time scales of the AIC and mirror instabilities are slower than the plasma advection time scales through these shocks. Consequently, the turbulence is built far from the shock front and the shock surface remains laminar. 
Small amplitude of magnetic field fluctuations in very weak shocks results from a low 
amount of free energy stored in temperature anisotropies at these shocks. 

\subsection{Dependence on the obliquity angle $\theta_{Bn}$}
\label{sec:angle}

\begin{figure}
	\centering\includegraphics[width=0.95\columnwidth, trim={0 1cm 0 1cm}, clip]{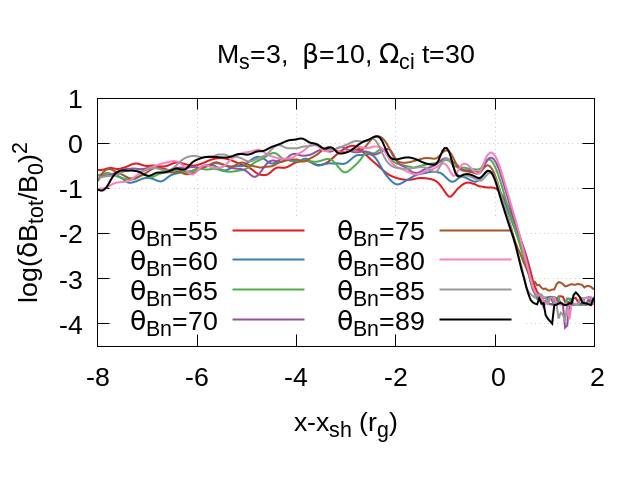}
     \caption{ The $y$-averaged profiles at time $\Omega_{ci}t=30$ of the normalized total energy density of the magnetic field fluctuations across the shocks with
$\beta=10$, $M_s=3$ and obliquity angles in the range $\theta_{Bn}=55^\circ-89^\circ$,
runs B1-B4, A3, and B5-B7 (compare Fig. \ref{fig:db_betas}).}
    \label{fig:db_angles}
\end{figure}

Figure~\ref{Fig:theta_comp} presents maps of the ion density and $B_z$ magnetic field fluctuations for shocks with $M_s=3$ in $\beta=20$ plasmas, and different $\theta_{Bn}$ angles, $55^\circ$, $75^\circ$, and $90^\circ$. 
Figure~\ref{fig:db_angles} shows profiles of the total energy density of the magnetic field turbulence for $M_s=3$ shocks in the range $\theta_{Bn}=55^\circ-89^\circ$ and for a lower value of plasma beta, $\beta=10$. 
We note, that all shocks are quasi-perpendicular, $\theta_{Bn}>45^\circ$, and all but the strictly-perpendicular one, $\theta_{Bn}=90^\circ$, are subluminal shocks with $\theta_{Bn} < \theta_{Bn, {\rm crit}}$, where the critical superluminality angle is $\theta_{Bn, {\rm crit}}=\arccos({u_{sh}/c})\approx 89.93^\circ$ for shocks in $\beta=20$ and $\theta_{Bn, {\rm crit}}\approx 89.95^\circ$ for $\beta=10$. 

The variations in the magnitude of the temperature anisotropy at the shock front with the field obliquity seen in Figure~\ref{Fig:comp_temp}c are related to the geometry of the ion reflection from the shock and their gyration in the upstream magnetic field. 
Upon reflection from the oblique shock, the ions acquire a finite drift in a direction parallel to the mean magnetic field that contributes to $T_\parallel$.
The apparent temperature anisotropy decreases as the parallel drift speed increases for smaller $\theta_{Bn}$.
An additional factor contributing to lower anisotropy at less oblique shocks is the ion drift in the motional electric field, the strength of which decreases with $\sin (\theta_{Bn})$. 

The patterns in the temperature anisotropy profiles observed with an increasing $\theta_{Bn}$ 
resemble those seen in temperature anisotropy profiles as Mach numbers progress from small to large values.
Consequently, more oblique shocks should generate faster-growing and stronger waves, whose amplitude peaks closer to the shock front, compared to shocks with smaller $\theta_{Bn}$. One can see such a trend in the $B_z$ maps in Figure~\ref{Fig:theta_comp} and in the magnetic energy density profiles in Figure~\ref{fig:db_angles}. 
However, despite a temperature anisotropy difference of approximately 3.5 between shocks with $\theta_{B_n}=55^\circ$ and $\theta_{B_n}=89^\circ$, there is only a slight difference, by a factor of a few, in the strength of the magnetic field fluctuations in these cases.
This might be because the free energy is provided not only by the temperature anisotropy but also by parallel ion drift. Consequently, a smaller anisotropy may not necessarily lead to a much lower level of turbulence.  

All shocks we analyze here develop density corrugations in the shock surface. Due to the slower growth of instabilities, the shock ripples emerge at later times at lower-$\theta_{Bn}$ shocks. 
The ripple wavelength becomes longer as the obliquity decreases. At the final simulation time $\Omega_{ci}t=56$ for $\beta=20$ shocks, Figure~\ref{Fig:theta_comp}, the ripple wave in the strictly-perpendicular shock, $\theta_{Bn}=90^\circ$, has a wavelength $\lambda_{\rm ripple}\approx 25\lambda_i$, in $\theta_{Bn}=75^\circ$ shock $\lambda_{\rm ripple}\approx 30\lambda_i$ (see also Section~\ref{sec:nonlinear}), and in the shock inclined at angle $\theta_{Bn}=55^\circ$ the ripple wavelength is $\lambda_{\rm ripple}\approx 50\lambda_i$. This dependence of $\lambda_{\rm ripple}$ on the field obliquity possibly results from the nonlinear evolution of the Alfvenic turbulence in the overshoot that couples the ion temperature anisotropy-driven modes with waves generated due to the parallel ion drift. Such dependence has not been reported before. We leave a detailed analysis of these effects to a future study.

We have demonstrated in Section~\ref{sec:nonlinear} that the wavelength of the ripple mode saturates past the simulation time $\Omega_{ci}t\approx 20$. We see a similar trend in the simulation of the strictly-perpendicular shock. As in the case with $\theta_{Bn}=75^\circ$ shock, the ripple wave structure is quasi-steady with smaller-scale density fluctuations occasionally appearing on top of the dominant mode. In this case, we also observe symmetric wave propagation parallel and anti-parallel to the mean magnetic field (see Section~\ref{sec:linear}). In the case of the oblique shock with $\theta_{Bn}=75^\circ$, the ripple wavelength evolves slowly. Confirming the saturation in this case necessitates an extended simulation time and possibly a larger computational box.

\section{Comparison with 3D simulations}
\label{App:3D}

Our studies in this work are mostly based on 2D simulations, since running 3D simulations encompassing large ion-scales of the shock-front ripples is computationally demanding, even in the hybrid approach. Therefore, to study the dimensionality effects, we performed two runs for shocks with Mach number $M_s=3$, obliquity angle $\theta_{Bn}=75^\circ$ and plasma beta $\beta=5$ and 10, runs E1-3D and E2-3D, respectively. As in 2D simulations, the upstream magnetic field lies in the $x-y$ plane. The size of the simulation box along the $y$ and $z$ directions is equal, $L_y=L_z$. All other simulation parameters are the same as in 2D runs.
The E1-3D run serves as a reference for studies of large-scale effects, while smaller-size E2-3D simulation provides supportive insights.

\begin{figure*}
\centering
\includegraphics[width=0.8\linewidth]{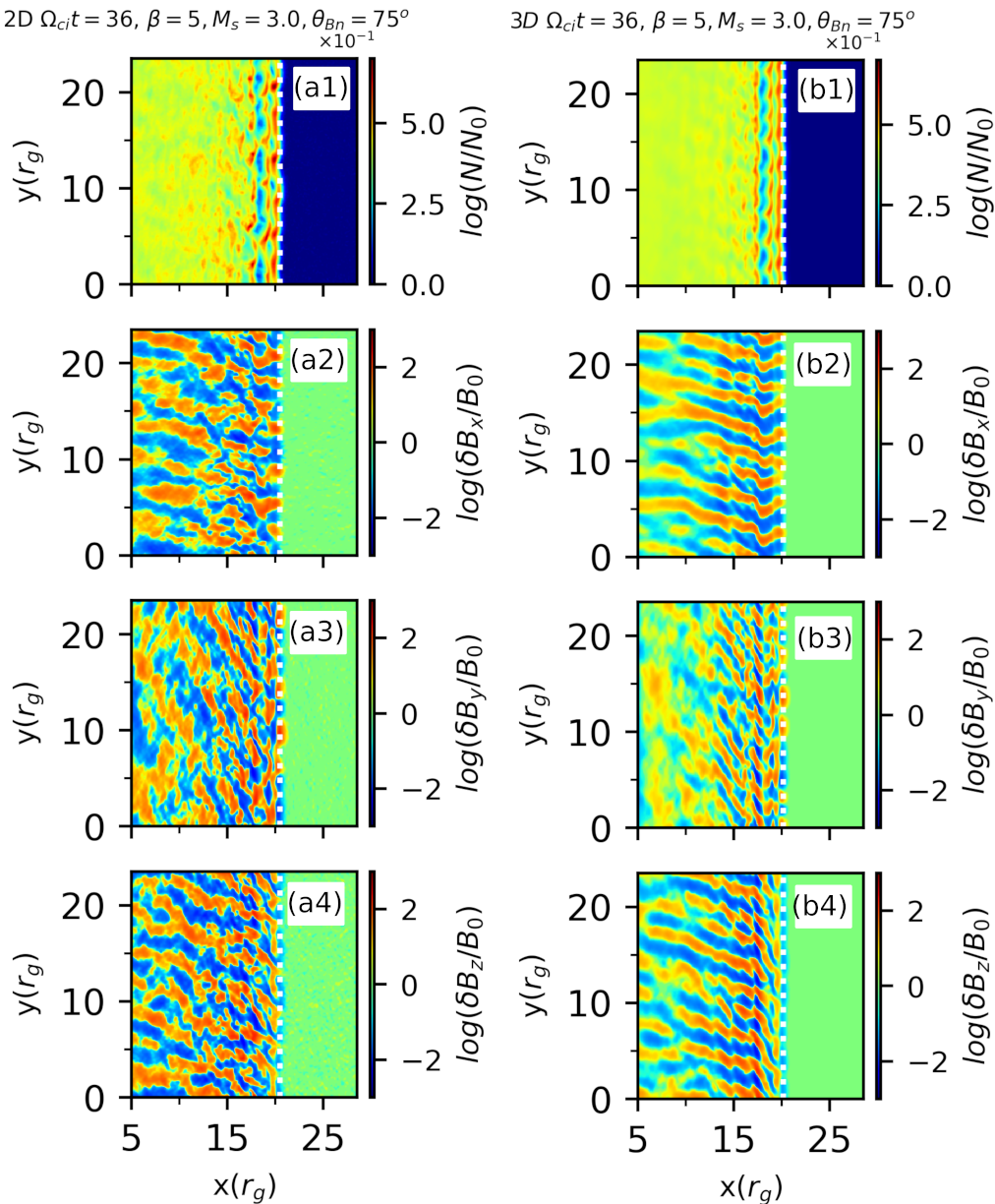}
    \caption{Structure of the $M_s=3$, $\beta=5, ~\theta_{Bn}=75^\circ$ shock at $\Omega_{ci}t=36$ in 2D and 3D simulation runs A2 (panels a*) and E1-3D (panels b*). Two-dimensional maps of the normalized ion density and magnetic field fluctuations are shown, and for run E1-3D the cut at $z=48 \lambda_i$ is displayed (see Fig.\ref{Fig:hb20_maps}). The shock position, $x_{sh}$, is marked with the white dotted line.}
    \label{fig:3D}
\end{figure*}

Figure~\ref{fig:3D} compares maps of the ion density and magnetic field fluctuations for the 2D run A2 (panels a*) and 
run E1-3D (panels b*) with $\beta=5$ at time $\Omega_{ci}t=36$. Maps for run E1-3D show a 2D cut across the computational box in the plasma flow direction 
in the middle of the box at $z=48\lambda_i$.  
One can note a very good correspondence in the density and wave mode structures between 2D and 3D. In particular, the wavelength $\lambda_{\rm ripple}\approx 12-14\lambda_i$ of the shock ripples in E1-3D run is in agreement with the ripple wavelength measured in 2D run A2 at the same simulation stage.

\begin{figure}
\centering
\includegraphics[width=\columnwidth]{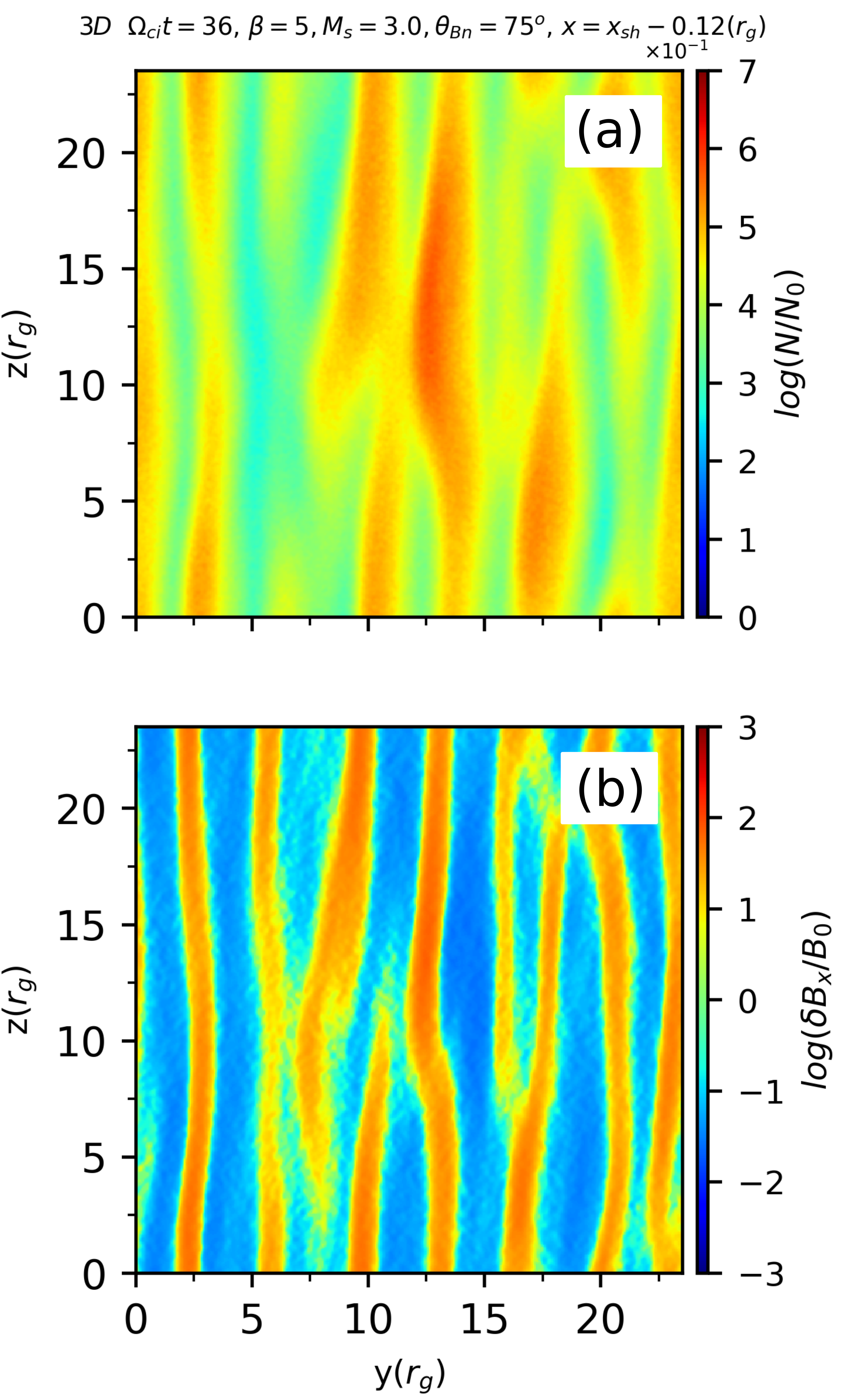}
    \caption{Maps of density and $B_x$ magnetic field fluctuations in the $y-z$ plane at $x=x_{sh}-0.12r_g$ (see Fig.~\ref{fig:3D}) at $\Omega_{ci}t=36$ for E1-3D run.}
    \label{fig:3Dxsh}
\end{figure}

The cut through the box in the direction transverse to the plasma flow at $x=x_{sh}-0.12 r_g$ presented in Figure~\ref{fig:3Dxsh}, shows the structure of the shock surface ripples and associated nonlinear AIC modes. It is apparent that the turbulence structure is 2D-like, and the waves have only weak long-wave components along the $z$-direction. 
We note that in the early-stage of the system evolution 
in the shock front region bounded by the first and the second overshoots,
there is a significant wave component to $\delta B_x, \delta B_y$ and $\delta B_z$ fluctuations with $k_z\approx k_y$, and the waves are oblique in general (not shown). However, a steady-state structure in this region is 2D-like, and only far downstream plasma hosts quasi-isotropic turbulence.
Similar characteristics are observed in run E2-3D with $\beta=10$.

\begin{figure}
\centering
\includegraphics[width=\columnwidth]{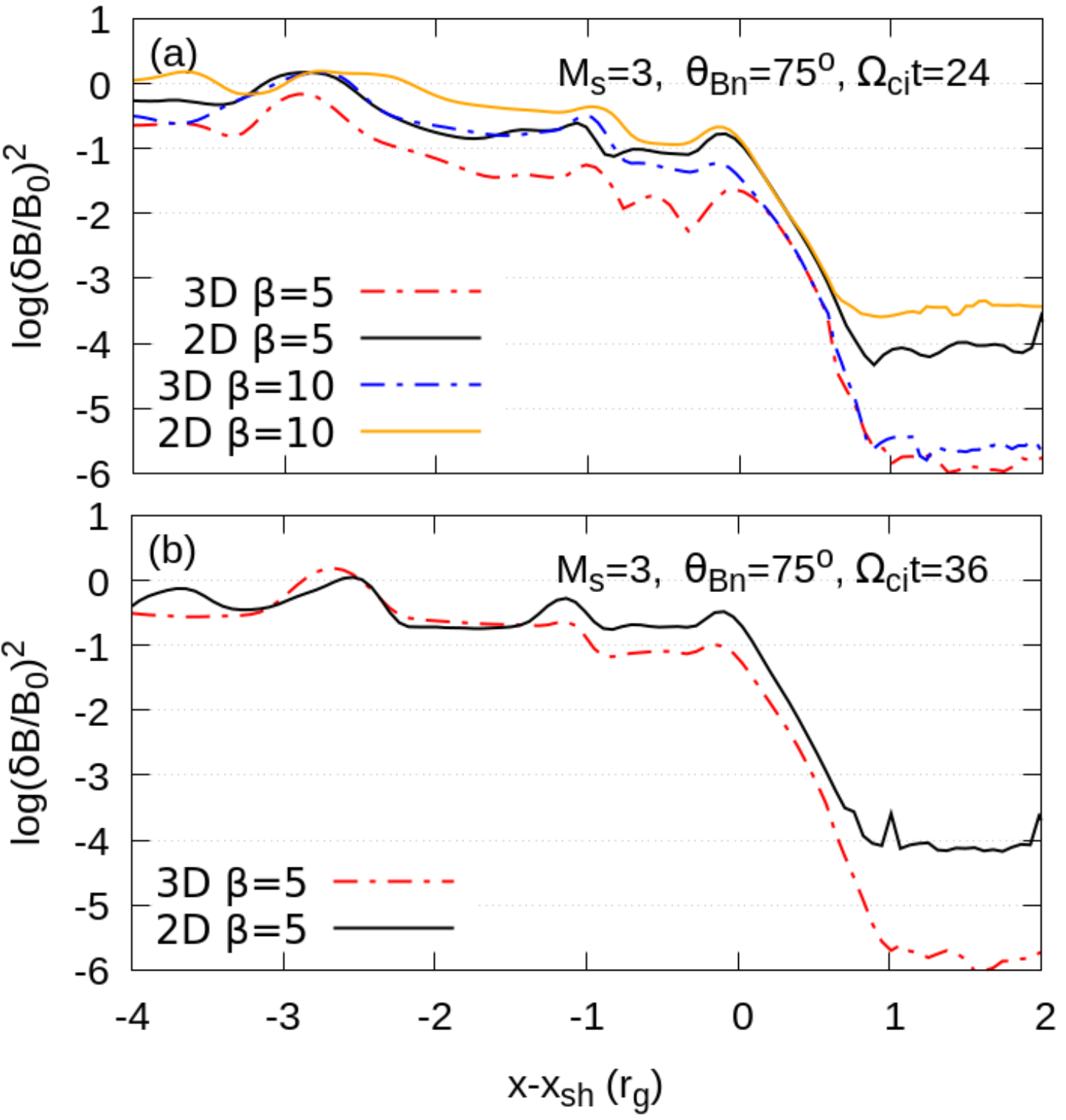}
    \caption{The $y$-averaged profiles of the normalized energy density of magnetic field fluctuations across shocks in 2D (solid lines) and 3D (dash-dotted lines) simulations, (a) at time $\Omega_{ci}t=24$ for shocks with $\beta=5$ and 10, and $M_s=3, ~\theta_{Bn}=75^\circ$, runs A2, A3, E1-3D, E2-3D, and (b) at $\Omega_{ci}t=36$ for runs A2 and E1-3D with $\beta=5$.}
    \label{fig:2D3D}
\end{figure}

Finally, Figures~\ref{fig:2D3D}a and \ref{fig:2D3D}b compare the profiles of the normalized energy density of the magnetic fluctuations between 2D (solid lines) and 3D (dash-dotted lines) simulations at $\Omega_{ci}t=24$ and $\Omega_{ci}t=36$, respectively. 
At an earlier stage, Figure~\ref{fig:2D3D}a, the magnetic field fluctuation levels in 2D and 3D are comparable in the case with $\beta=10$. At the same time, shocks with $\beta=5$ differ between 2D and 3D by nearly an order of magnitude in the average magnetic energy density in the region close to the shock front. Similar magnetic turbulence levels are nevertheless reached by $\Omega_{ci}t=36$ in this case, Figure~\ref{fig:2D3D}b, because the wave growth rates are lower for smaller $\beta=5$, compared to $\beta=10$ (see Section~\ref{sec:beta}).

The results presented in this section indicate that the steady-state structure of weak shocks in high-beta plasmas does not considerably differ between fully 3D simulations and simulations with dimensionality restricted to 2D with an in-plane magnetic field geometry. 
While this conjecture may not hold for strong shocks, like the case with $M_s=5.5$ considered here in 2D (see Section~\ref{sec:mach}), it is expected to be valid for shocks with $M_s\lesssim 3$, that are considered more representative of the merger shocks in ICM.
Further studies are necessary to demonstrate conclusively that the effects of higher plasma beta or different magnetic field obliquity do not alter this conclusion.

\section{Summary and concluding remarks}
\label{sec:4}

In this study we employ hybrid-kinetic numerical simulations to investigate the ion-scale kinetic instabilities at merger shocks of galaxy clusters.
Our results can be summarized as follows:
\begin{itemize}
    \item[$\bullet$] The structure of a subluminal large-scale 2D shock with fiducial merger shock parameters, $M_s=3$, $\beta=20$ and $\theta_{Bn}=75^\circ$, is governed by the ion temperature anisotropies in the shock due to the ion reflection from the shock and their gyration in the magnetic field in the shock transition. These anisotropies drive multi-scale magnetic turbulence whose properties in the linear stage are consistent with AIC and mirror modes and agree very well with characteristics observed in fully kinetic PIC simulations. This confirms the credibility of hybrid simulations in exploring the ion-scale physics of shocks.
    \item[$\bullet$] With growing plasma $\beta$ shocks generate stronger turbulence that is confined closer to the shock front. Shocks in very high-beta plasmas, $\beta\approx 50-100$, produce an order of magnitude more intense turbulence than lower-$\beta$ shocks.
    \item[$\bullet$] Nonlinear evolution of AIC and mirror modes, and the nonlinear couplings between these modes are responsible for the formation of plasma density fluctuations in the shock front. Initial short-wave ripples evolve toward long wavelengths. We show for the first time that in $M_s=3$ shocks in high-$\beta$ plasmas, $\beta=5-50$, the growth of the shock front corrugations and magnetic waves saturates. The shock ripples reach wavelengths of  $\lambda_{\text{ripple}} \approx 3.7-3.9 r_g$.
    This result has important consequences for electron acceleration processes at cluster shocks. The saturated wavelengths of the waves impose constraints on the maximum energy of particles that are confined at the shock through resonant scattering off the waves and undergo the SSDA process.
    \item[$\bullet$] At very weak shocks with $M_s\lesssim 2.3$ the AIC growth time scale is slower than the plasma advection time scale through the shock. Therefore,
    such shocks do not develop shock surface ripples and form weak turbulence in the shock downstream.
    This is consistent with the notion of the critical Mach number $M_{\rm AIC}^*\approx 2.3$ to trigger multi-scale waves in the shock front, proposed in \cite{Ha2021}.
    \item[$\bullet$] For the first time we also investigate high-$\beta$ quasi-perpendicular shocks in a wide range of the magnetic field obliquity. Despite significant variations in temperature anisotropy at the shock, we only observe a moderate differences in the strength of magnetic field fluctuations at various $\theta_{Bn}$. This indicates that, in addition to temperature anisotropy, parallel ion drift plays a significant role in driving magnetic turbulence. We also observe that shock ripples have longer wavelengths as the field obliquity decreases. Our results indicate that the ripple growth saturation occurs independent of $\theta_{Bn}$. However, to confirm this conjecture, longer simulation times and potentially larger computational boxes are required for shocks with low $\theta_{Bn}$.     
    \item[$\bullet$] The comparison between 2D and 3D simulations suggests that the steady-state structure of merger shocks with \(M_s\lesssim 3\) in high-beta plasmas does not significantly differ between fully 3D and 2D simulations with an in-plane mean magnetic field. The shock surface ripples and associated nonlinear Alfven ion cyclotron (AIC) modes exhibit a 2D-like structure at the shock front and immediate downstream, with quasi-isotropic turbulence emerging only in the far downstream region. Steady-state shocks in both 2D and 3D develop comparable levels of turbulence. 
    
\end{itemize}

The studies in this work provide the fundamental framework for conducting comprehensive large-scale 3D simulations aimed at understanding shock physics in a wide range of physical conditions. They may also guide parameter choice for large-scale PIC simulations. Equally important is that they lay the foundation for further investigations of the impact of ion-scale turbulence on electron acceleration in high-beta shocks by means of test-particle simulations \citep[e.g.,][]{TB19} or similar methods. 
As emphasized in Section~\ref{sec:intro}, 
multi-scale turbulence is vital for efficient electron acceleration via the SSDA process. Our recent PIC simulations \citep{Kobzar21} indicate that electron-scale turbulence has less impact on electron energization to supra-thermal energies. 
Hence, the investigation of electron acceleration can be reliably conducted through test particle simulations. Our generalized fluid particle hybrid code is particularly well-suited for this purpose, as it allows for the incorporation of additional energetic electron component and the exploration of electron acceleration within turbulent shock structures. Such investigations are currently ongoing.

\begin{acknowledgements} We thank the anonymous referee for useful comments. The authors acknowledge the support of Narodowe Centrum Nauki through research projects no. 2019/33/B/ST9/02569 (S.B. and J.N.) and no. 2016/22/E/ST9/00061 (O.K.). We gratefully acknowledge Polish high-performance computing infrastructure PLGrid (HPC Centers: ACK Cyfronet AGH) for providing computer facilities and support within computational grant no. PLG/2022/015967.
This research was supported by the International
Space Science Institute (ISSI) in Bern, through ISSI International
Team project \# 520 (Energy Partition across Collisionless Shocks).
\end{acknowledgements}


\bibliographystyle{aa}
\bibliography{bib} 

\begin{appendix}

\section{Mass ratio dependence}\label{Ap:mass}
The hybrid simulation code adopted in this study includes the finite electron inertia effect. The model thus correctly considers the dispersive effect on the linear dispersion relation associated with the finite electron mass, which appears at the electron skin depth scale. This indicates that the finite mass ratio should not affect the results as long as the electron skin depth is unresolved. In this case, we can rather consider the finite inertia effect as a numerical stabilization trick to suppress high-frequency whistler noise \citep{Amano2014}.

\begin{figure}[h!]
\centering
	\includegraphics[width=0.9\columnwidth]{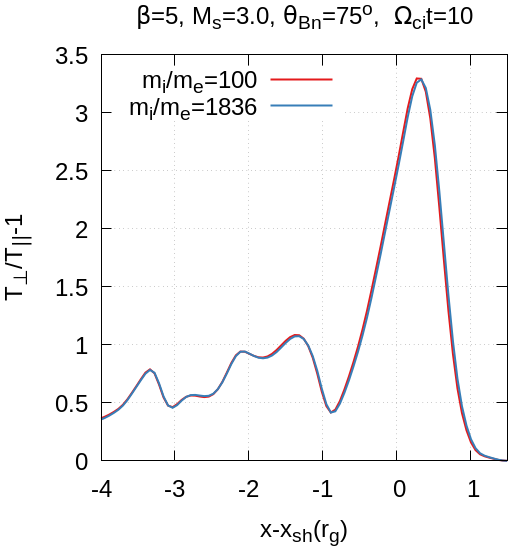}
    \caption{Temperature anisotropy profiles across the shock transition at $\Omega_{ci}t=10$, showing a dependence on the ion-to-electron mass ratio, $m_i/m_e$, for the test case of $\beta=5$, $M_s=3.0$, $\theta_{B_n}=75^o$.}
    \label{fig:mime}
\end{figure}

As an example, Fig.~\ref{fig:mime} compares the snapshots of the temperature anisotropy at $\Omega_{ci} t = 10$ for $\beta = 5$, $M_s = 3.0$, $\theta_{B_n} = 75^o$ with two different mass ratios $m_i/m_e = 100$ and $1836$. As we can see, the results are almost identical. This is natural since, in both cases, the spatial resolution $0.25 \lambda_i$ is larger than the electron skin depth ($0.1 \lambda_i$ and $0.02 \lambda_i$ for $m_i/m_e = 100$ and $1836$, respectively).

Note that, in general, attempting to resolve the electron skin depth with the hybrid code is not advisable.
In fact, the electron kinetic effect, such as the cyclotron damping ignored in the model, starts to play a role at this scale. Therefore, one has to be careful in interpreting the physics at the electron scale in the hybrid model.

\section{Dependence on the number of particles per cell}\label{Ap:npc} 
\begin{figure}
\centering
	\includegraphics[width=0.9\columnwidth]{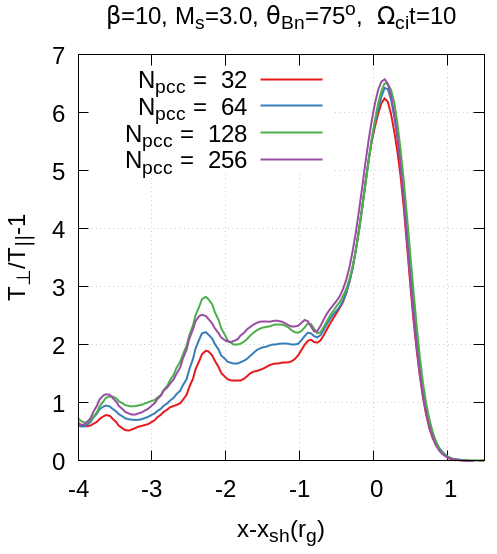}
    \caption{Temperature anisotropy profiles across the shock transition at $\Omega_{ci}t=10$, showing a dependence on the number of particles per cell, $N_{ppc}$, for the test case of $\beta=10$, $M_s=3.0$, $\theta_{B_n}=75^o$.}
    \label{fig:npc}
\end{figure}

In Fig. \ref{fig:npc}, we explore the impact of varying the number of computational particles per cell, $N_{ppc}$, with values of $N_{ppc}=32$ (red), 64 (blue), 128 (green), and 256 (purple). Noticeable distinctions are not particularly evident within the shock front, where the temperature anisotropy is most pronounced, especially for $N_{ppc}>64$. However, these differences become noteworthy past the shock front, where strong magnetic turbulence develops. For $N_{ppc}>128$ the results are convergent. We use $N_{ppc}=128$ in our simulations.

\section{Periodic box simulation of the wave turbulence evolution}\label{Ap:periodic}

In Figure~\ref{Fig:fourier_t10} in Section~\ref{sec:linear}, we present the results of the linear analysis using the local macroscopic properties of the plasma at two regions of the shock transition -- the shock ramp and the region between the undershoot and the second overshoot. 
To confirm the compatibility of early-stage instabilities with linear calculations and explore the nonlinear evolution of these instabilities, we conducted 2D simulations with periodic boundary conditions for the values at the shock ramp. Hence, we assume the temperature anisotropy $T_{\perp}/T_{\parallel}=7$, parallel ion beta $\beta_{\parallel}=9.33$, and the electron plasma beta  $\beta_e=10$.   
Note that while these parameters are measured in the simulation run A4 at time \(\Omega_{ci}t=10\), they show no significant differences from parameters measured at the shock ramp at earlier times. Therefore, they should be considered representative of the early-stage shock evolution at this location.

The simulation domain is a square box of size $L_x = L_z = 60\lambda_i^\prime$ in the $x-y$ plane, and the mean magnetic field $\mathbf{B_0}$ is along the $x$-direction (note that in shock simulations, the compressed downstream magnetic field is essentially aligned with the $y$-direction). Here, we denote the unit length scale $\lambda_i^\prime$ to emphasize that wavelengths and wave vectors are given in terms of the local values measured in the shock simulation A4. All other simulation parameters are the same as in the shock simulations. 

\begin{figure}
	\includegraphics[width=\columnwidth]{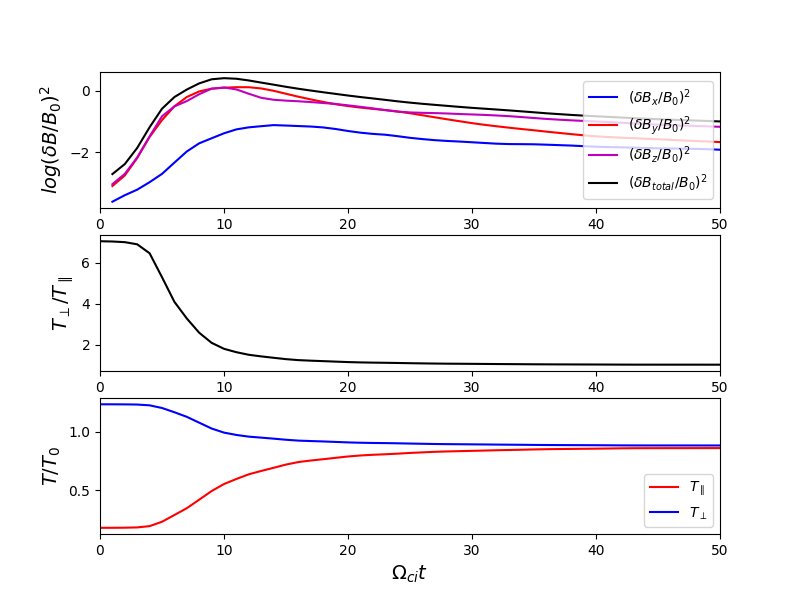}
    \caption{Time evolution of the magnetic field fluctuation energy densities (top), the temperature anisotropy (center), and the parallel and perpendicular temperature components (bottom) for the periodic-box simulation.}
    \label{fig:periodic_energy}
\end{figure}

For the given simulation setup, the early evolution of the instabilities should correspond to the linear theory results presented in Figure~\ref{Fig:fourier_t10}a. We expect that the AIC waves are formed in $B_y$ and $B_z$ magnetic field components, and the oblique compressive mirror mode is seen primarily in the $B_x$ component. 
Figure~\ref{fig:periodic_energy} shows the time evolution of the magnetic field fluctuations and temperature profiles. Field turbulence rapidly grows and saturates around $\Omega_{ci}t\approx 10$, coinciding with the near disappearance of temperature anisotropy. Subsequently, the system further relaxes toward complete ion distribution isotropy. In this regard, periodic box simulation results significantly differ from the shock simulations. In the latter, a continuous inflow of fresh ions from the upstream sustains high-temperature anisotropy at the shock, leading to the excitation of instabilities. Consequently, meaningful comparisons between the two simulations are viable only during the early evolution of the periodic-box system, for $\Omega_{ci}t\lesssim 10$. 

In Figure~\ref{fig:periodic_energy}, it is evident that the growth rate of transverse magnetic field fluctuations, $\delta B_y$ and $\delta B_z$, associated primarily with AIC modes, significantly exceeds that of $\delta B_x$ waves, corresponding to mirror modes. During the time interval $\Omega_{ci}^\prime t=2.5-5$, the measured growth rates are $\gamma_{\rm AIC}\approx 0.8\Omega_{ci}^\prime$ for $\delta B_y, \delta B_z$ components and $\gamma_m\approx 0.4\Omega_{ci}^\prime$ for $\delta B_x$. The slower growth of mirror modes aligns with linear theory. However, both growth rates are notably smaller than the linear predictions of $\gamma_{\rm AIC}\approx 1.2\Omega_{ci}^\prime$ and $\gamma_m\approx 0.6\Omega_{ci}^\prime$. This suggests that, even at this early stage, nonlinear effects are at play, which is unsurprising given the challenge of capturing a truly linear evolution when the wave growth rate is very high. Nonetheless, we believe comparing early-stage periodic-box turbulence with its characteristics observed at the shock ramp is viable.  

\begin{figure*}[]
\centering
 \includegraphics[width=1.0\linewidth]{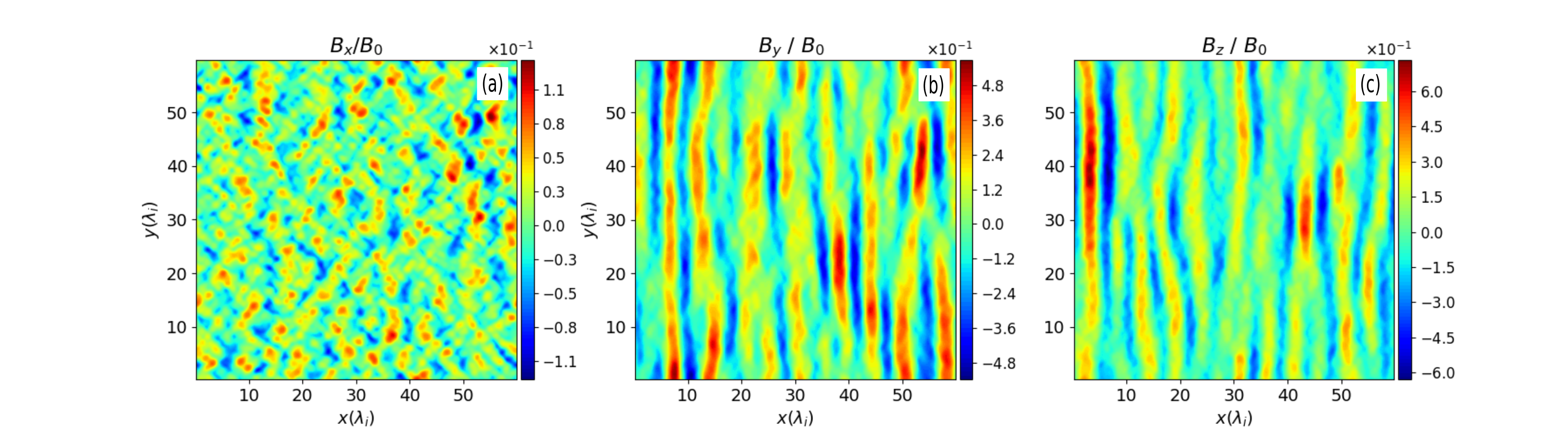}\\ 
 \caption{Maps of the normalized magnetic field fluctuations for the periodic box simulation at time $\Omega_{ci}^\prime t=4$. From left to right shown are (a) $\delta B_x=B_x-B_0$, (b) $\delta B_y$, and (c) $\delta B_z$.}
 \label{fig:periodic_maps}
\end{figure*}

\begin{figure*}[]
\centering
 \includegraphics[width=\linewidth]{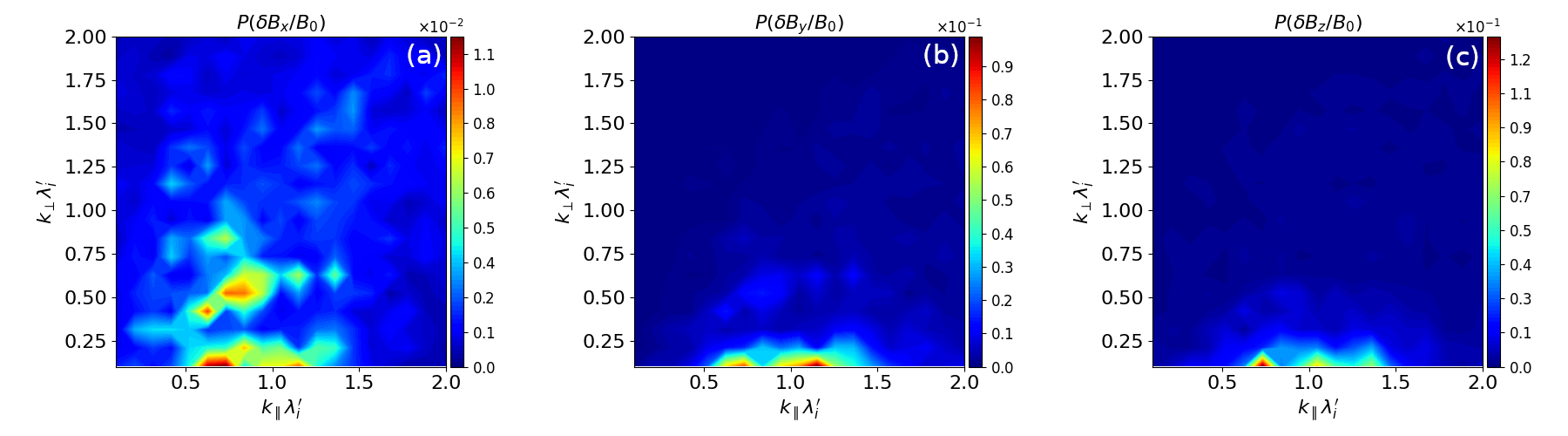}
 \caption{Fourier power spectra of magnetic field fluctuations in the periodic box simulation averaged in the time interval  $3\leq\Omega_{ci}^\prime t\leq5$. Note that the dynamic range is different in each panel.}
 \label{fig:periodic_fourier}
\end{figure*}

Figure~\ref{fig:periodic_maps} shows distributions of $\delta B_x$, $\delta B_y$, and $\delta B_z$ magnetic field fluctuations at time $\Omega_{ci}^\prime t=4$. Figure~\ref{fig:periodic_fourier} displays corresponding Fourier power spectra averaged in the time interval $3 \le \Omega_{ci}^\prime t\le 5$.  
The Fourier power in $B_y$ and $B_z$ waves is maximum at $k_\parallel\lambda_i^\prime\approx 0.7-1.4$, in reasonable agreement with linear theory (compare Figure~\ref{Fig:fourier_t10}a). The corresponding range of wavelengths calculated using the shock simulation values is $\lambda\approx 3.7-7.3\lambda_i$, which is consistent with the AIC waves observed at the first overshoot (compare Figure~\ref{Fig:fourier_t10}a and~\ref{Fig:fourier_t10}c).
Slightly oblique waves in $B_x$ with wave vectors $k_\perp\lambda_i^\prime\lesssim 0.25$ and similar $k_\parallel\lambda_i^\prime\approx 0.6-1.4$ have also their counterparts in the waves observed at overshoot in $\delta B_y$ fluctuations (see description in Section~\ref{sec:linear}). Although these waves are most likely due to nonlinear effect of AIC and mirror waves coupling, the oblique modes with $k_\parallel\lambda_i^\prime\approx 0.6-1.4$ and $k_\perp\lambda_i^\prime\approx 0.4-0.7$ seen in $\delta B_x$ and also $\delta B_y$ are consistent with the predictions of the linear analysis.
We conclude that periodic-box simulations satisfactorily reproduce both the linear theory predictions and the properties of waves observed in the first overshoot in the early-stage evolution of the shock.

During subsequent evolution, we observed waves in all magnetic field components growing in wavelengths (not shown). This is in line with the known nonlinear evolution of AIC and mirror waves \citep[e.g.,][]{Shoji2009}.


\section{Comparison with 2D PIC simulations}\label{Ap:PIC_hybrid}

\begin{center}
\begin{figure*}[ht]
\centering
  \subfloat[$M_s=3, \theta_{Bn}=75^{\circ}, \beta=5$ (PIC)]
 {\includegraphics[width=0.332\linewidth]{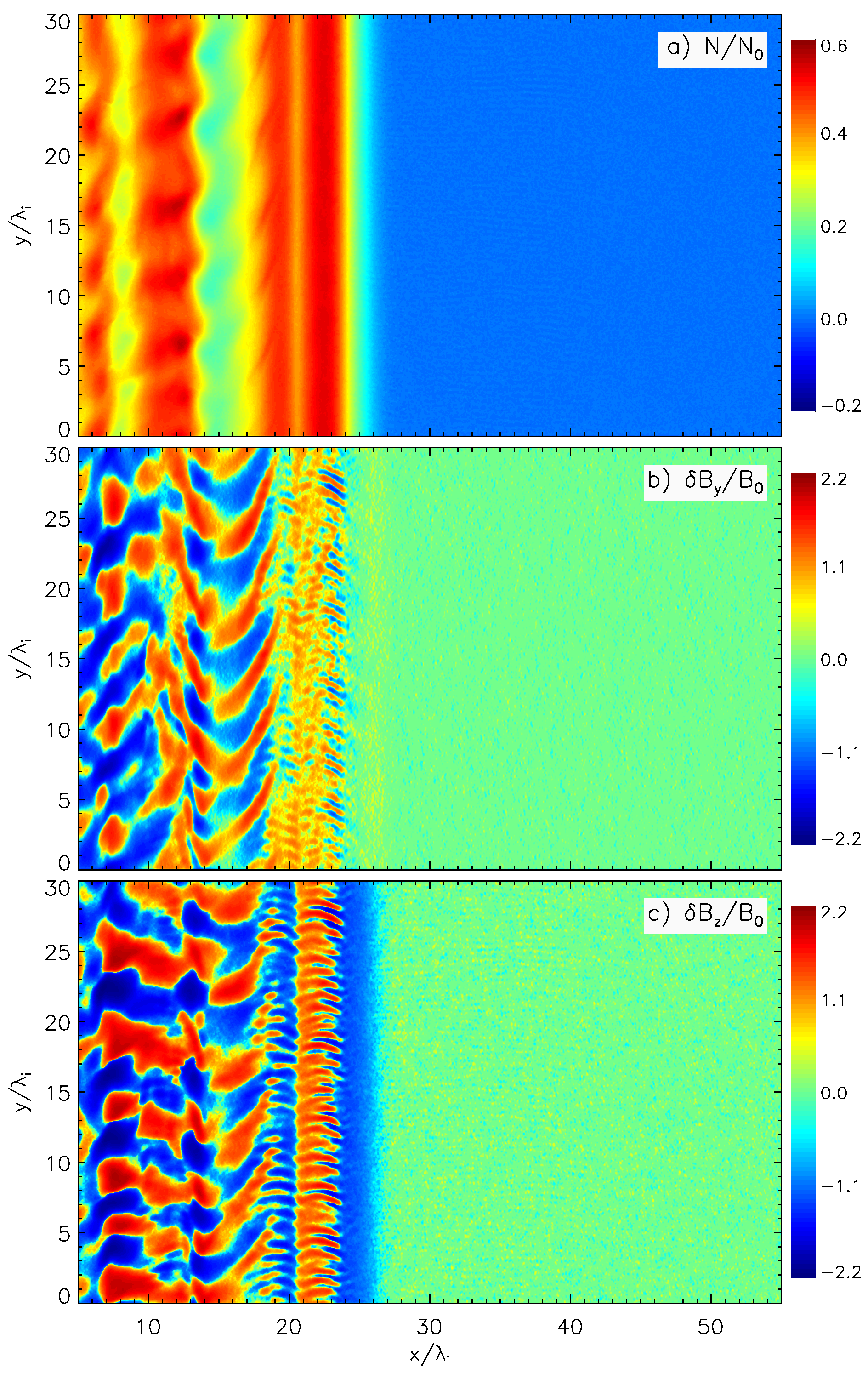}}
  \subfloat[$M_s=3, \theta_{Bn}=75^{\circ}, \beta=5$ (hybrid)]
 {\includegraphics[width=0.3495\linewidth]{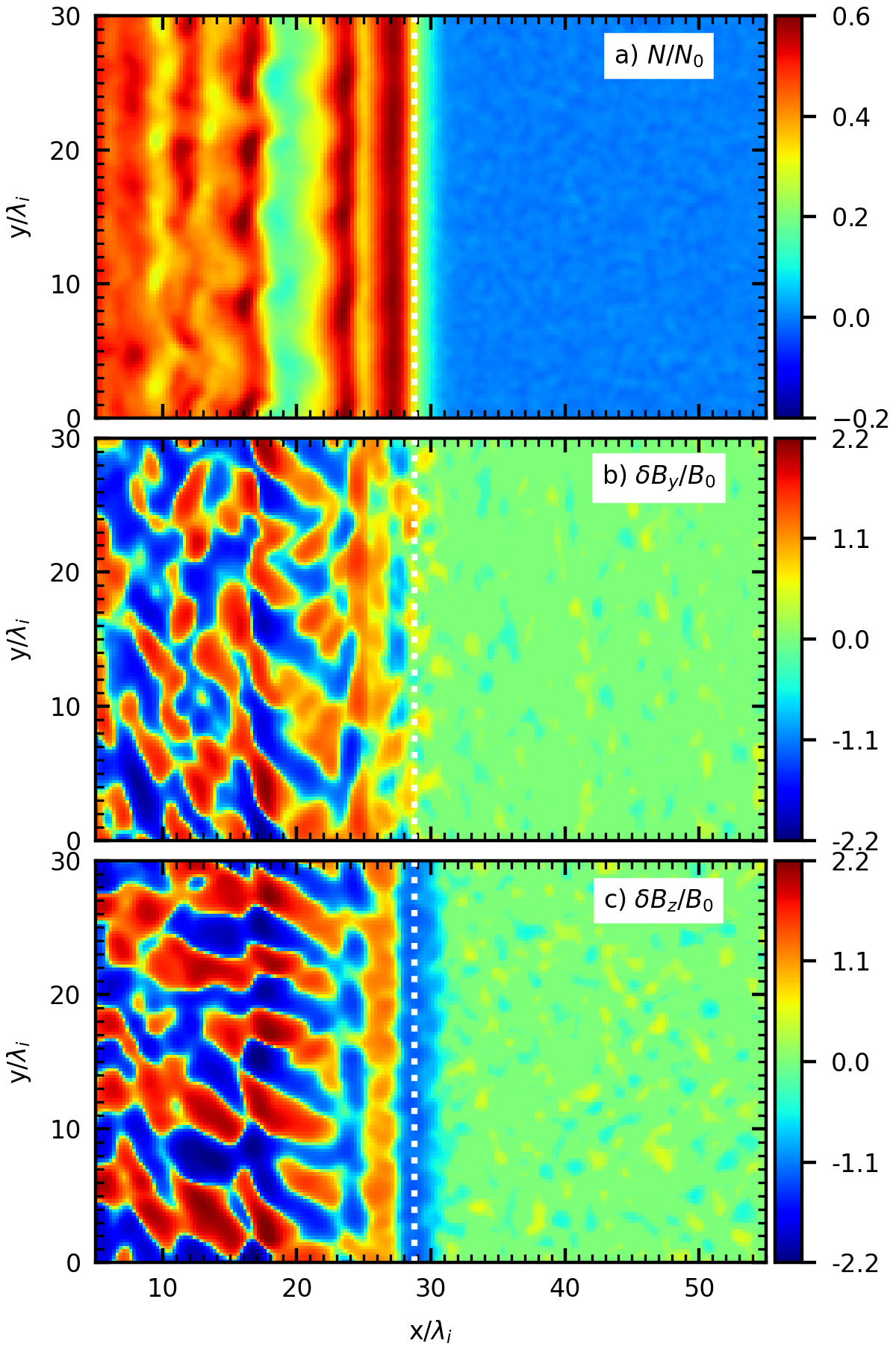}}\hfill\\
  \subfloat[$M_s=3, \theta_{Bn}=75^{\circ}, \beta=20$ (PIC)]
  {\includegraphics[width=0.332\linewidth]{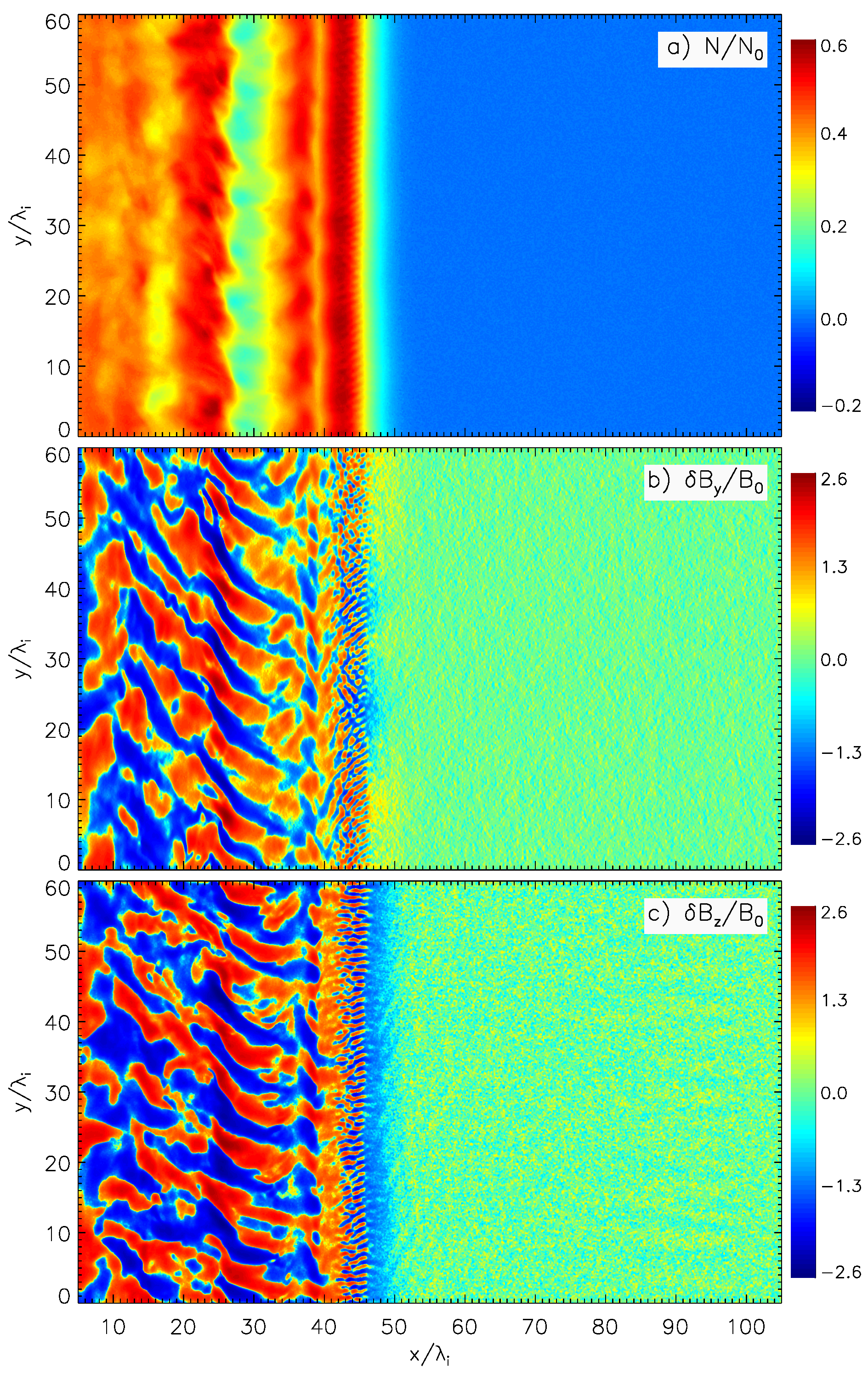}}
  \subfloat[$M_s=3, \theta_{Bn}=75^{\circ}, \beta=20$ (hybrid)]
  {\includegraphics[width=0.3495\linewidth]{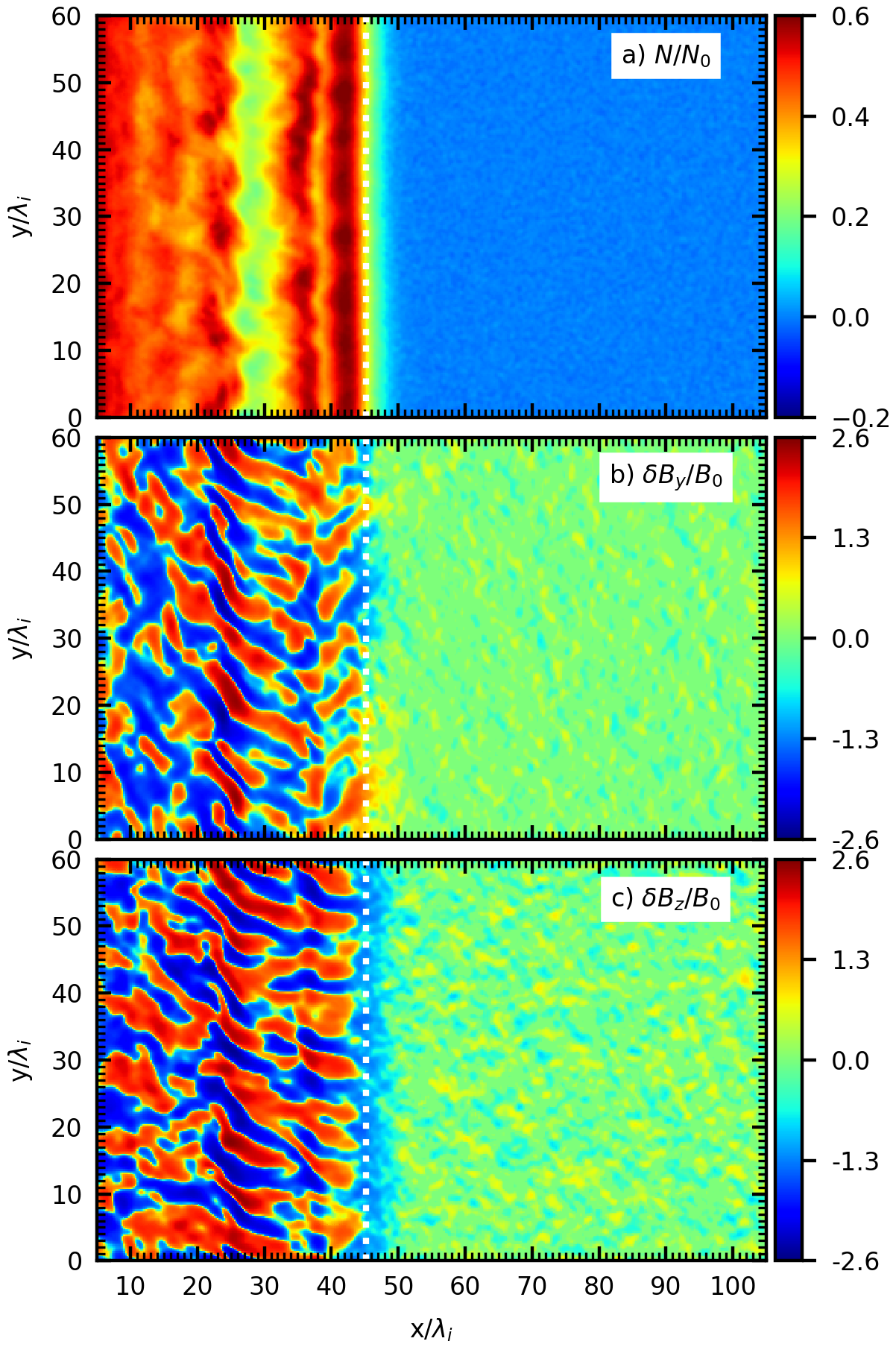}}\hfill  
  \caption{Comparison of the shock structures obtained with PIC (\emph{left} panels) and hybrid (\emph{right} panels) simulations for shocks with $M_s=3$ and $\theta_{Bn}=75^{\circ}$.
  \emph{Top} row shows results for $\beta=5$ at time $\Omega_{ci}t=10$ for the PIC (a) and $\Omega_{ci}t=13$ for the  hybrid (b) simulation (run A2). \emph{Bottom} row displays results 
  for $\beta=20$ at time $\Omega_{ci}t=10$ for the PIC (c) and $\Omega_{ ci}t=11$ for the  hybrid (d) simulation (run A4). In each panel, from top to bottom, normalized distributions of ion density, and $\delta B_y$ and $\delta B_z$ magnetic field fluctuations are shown. Note, that scaling of all distributions is logarithmic (see Fig.~\ref{Fig:hb20_maps}). Only portions of the computational boxes are shown for hybrid simulation runs A2 and A4.}
  \label{Fig:hybridpicb5b20}
\end{figure*}
\end{center}

In Figure~\ref{Fig:hybridpicb5b20}, we present a comparison of 2D results obtained for the same shock parameters using two different simulation approaches: particle-in-cell (PIC) simulations in the left panels (a) and (c) and hybrid simulations in the right panels (b) and (d)). We compare the early stage structures of shocks with Mach number $M_s=3$, mean magnetic field inclination of $\theta_{Bn}=75^{\circ}$, and two different plasma betas: $\beta=5$ in top panels and $\beta=20$ in bottom panels. 
PIC simulations results for the case of $\beta=5$ are taken from~\cite{Kobzar21}. The results for $\beta=20$ have not yet been published (Kobzar et al. (2024), in preparation). We present hybrid simulations results from runs A2 and A4, though for a clearer one-to-one comparison, we display only portions of the computational boxes for hybrid runs in Figure~\ref{Fig:hybridpicb5b20}.
The shock structures are compared at slightly different times due to the distinct nature of the two simulation codes. Nevertheless, the results are considered to be contemporaneous.
PIC simulations have been performed with 20 particles per cell per species, ion-to-electron mass ratio $m_i/m_e=100$, and the electron skin depth  $\lambda_e=c/\omega_{pe}=15$ cells, where $\omega_{pe}=\sqrt{4\pi e^2N_e/m_e}$ is the electron plasma frequency.

In the linear phase of PIC simulations, the prevailing turbulence in the first overshoot is the electron whistler waves, evident as short-scale magnetic field fluctuations at $18\lesssim x/\lambda_i\lesssim 24$ in Figure~\ref{Fig:hybridpicb5b20}a and at $41\lesssim x/\lambda_i\lesssim 46$ in Figure~\ref{Fig:hybridpicb5b20}c (middle and bottom panels). 
However, strong ion-scale fluctuations grow in parallel and emerge downstream of the first overshoot. The characteristics of these waves in $\delta B_x$ (not shown) and $\delta B_z$ components are consistent with AIC instability modes, and in the $\delta B_y$ component with the mirror modes.   
The structure of these fluctuations closely resembles the waves observed in hybrid simulations in the same shock regions. In particular, in the $\beta=5$ shock, the dominant wavelength of the nearly transverse AIC fluctuations is approximately $\lambda_{\rm AIC}\approx 5\lambda_i$, and the field-aligned component of the oblique mirror modes, $\lambda_{m\parallel}\approx 6\lambda_i$, in both cases. Similarly, the wavelengths of both the AIC and mirror waves are $\lambda_{\rm AIC}\approx\lambda_{m\parallel}\approx 7.5\lambda_i$ in the shocks in $\beta=20$ plasmas.   

As the PIC and hybrid simulations progress, the ion-scale waves evolve into multi-scale turbulence. This includes the development of long-wave shock-front rippling modes,  whose nonlinear wavelengths closely match. For instance, the wavelength of $\lambda_{\text{ripple}} \approx 16\lambda_i$ of the ripples observed at time $\Omega_{ci}t = 36$ in run A2, closely fits the shock corrugation wavelength reported in \cite{Kobzar21}. 
The agreement between the results obtained with the hybrid-kinetic approach and our PIC simulations, as well as other PIC simulation studies~\citep[e.g.,][]{Ha2021,Guo_2017,Guo2019} attests to the validity of the hybrid simulations in exploring ion-scale shock physics and the role of ion-scale turbulence in particle acceleration. 

\end{appendix}

\end{document}